\documentclass[twocolumn,times]{aastex62}
\usepackage{natbib}
\usepackage{graphicx}
\usepackage[space]{grffile}
\usepackage{latexsym}
\usepackage{amsfonts,amsmath,amssymb}
\usepackage{url}
\usepackage[utf8]{inputenc}
\usepackage{fancyref}

\received{December 12, 2017}
\revised{February 14, 2018}
\accepted{\today}

\shorttitle{The Magnetic Response to Umbral Flashes}
\shortauthors{Houston et al.}
\newcommand{\CaIR}{{Ca~{\sc{ii}} 8542{\,}{\AA}}}
\newcommand{\HeI}{{He~{\sc{i}} 10830{\,}{\AA}}}
\newcommand{\strongarrow}{\mathbin{\rotatebox[origin=c]{75}{$\leftrightarrow$}}}
\newcommand{\weakarrow}{\mathbin{\rotatebox[origin=c]{-15}{$\leftrightarrow$}}}

\begin{document}
\title{The Magnetic Response of the Solar Atmosphere to Umbral Flashes}
\correspondingauthor{S.~J. Houston}
\email{shouston22@qub.ac.uk}

\author[0000-0001-5547-4893]{S.~J. Houston}
\affiliation{Astrophysics Research Centre, School of Mathematics and Physics, Queen's University Belfast, Belfast, BT7 1NN, UK}                                  
\author[0000-0002-9155-8039]{D.~B. Jess} 
\affiliation{Astrophysics Research Centre, School of Mathematics and Physics, Queen's University Belfast, Belfast, BT7 1NN, UK}                                  
\affiliation{Department of Physics and Astronomy, California State University Northridge, Northridge, CA 91330, U.S.A.}      
\author[0000-0002-1248-0553]{A. Asensio Ramos}
\affiliation{Instituto de Astrof\'isica de Canarias, C/V\'ia Lactea s/n, E-38205 La Laguna, Tenerife, Spain}
\affiliation{Departamento de Astrof\'isica, Universidad de La Laguna, E-38206 La Laguna, Tenerife, Spain}   
\author[0000-0001-5170-9747]{S.~D.~T. Grant}
\affiliation{Astrophysics Research Centre, School of Mathematics and Physics, Queen's University Belfast, Belfast, BT7 1NN, UK}
\author{C. Beck}
\affiliation{National Solar Observatory (NSO), Boulder, CO 80303, U.S.A.}
\author[0000-0003-2622-7310]{A.~A. Norton}
\affiliation{Hansen Experimental Physics Laboratory (HEPL), Stanford University, Stanford, CA 94305, U.S.A}
\author{S. Krishna Prasad} 
\affiliation{Astrophysics Research Centre, School of Mathematics and Physics, Queen's University Belfast, Belfast, BT7 1NN, UK}     

\begin{abstract}
Chromospheric observations of sunspot umbrae offer an exceptional view of magneto-acoustic shock phenomena and the impact they have on the surrounding magnetically-dominated plasma. We employ simultaneous slit-based spectro-polarimetry and spectral imaging observations of the chromospheric {\HeI} and {\CaIR} lines to examine fluctuations in the umbral magnetic field caused by the steepening of magneto-acoustic waves into umbral flashes. Following the application of modern inversion routines, we find evidence to support the scenario that umbral shock events cause expansion of the embedded magnetic field lines due to the increased adiabatic pressure. The large number statistics employed allow us to calculate the adiabatic index, $\gamma = 1.12\pm0.01$, for chromospheric umbral locations. Examination of the vector magnetic field fluctuations perpendicular to the solar normal revealed changes up to $\sim$200~G at the locations of umbral flashes. Such transversal magnetic field fluctuations have not been described before. Through comparisons with non-linear force-free field extrapolations, we find that the perturbations of the transverse field components are orientated in the same direction as the quiescent field geometries. This implies that magnetic field enhancements produced by umbral flashes are directed along the motion path of the developing shock, hence producing relatively small changes, up to a maximum of $\sim$8 degrees, in the inclination and/or azimuthal directions of the magnetic field. Importantly, this work highlights that umbral flashes are able to modify the full vector magnetic field, with the detection of the weaker transverse magnetic field components made possible by high-resolution data combined with modern inversion routines.
\end{abstract}

\keywords{shock waves --- Sun: chromosphere --- Sun: magnetic~fields --- Sun: oscillations --- Sun: photosphere --- sunspots}

\section{Introduction}{\label{sec:Intro}}
The study of magnetic field fluctuations and oscillations in the solar atmosphere is in its relative infancy \citep{2002AN....323..317S}. To date, the majority of studies have focused on photospheric oscillations to determine how magneto-hydrodynamic (MHD) waves propagate and subsequently channel energy into the chromosphere and corona \citep[e.g.,][]{2000ApJ...534..989B, 2009ApJ...702.1443F,2015LRSP...12....6K,2016NatPh..12..179J,2016ApJ...831...24K}. Similar studies in the more-diffuse chromosphere have proved difficult to undertake for a number of reasons, namely the advanced observational instrumentation required \citep{2015SSRv..190..103J}, in conjunction with the necessary development needed for inversion techniques to be able to infer plasma parameters within the complex physical conditions of the chromosphere \citep{2016LRSP...13....4D}. 

There are a few studies that have started to provide insight into chromospheric magnetic field perturbations, through the analysis of non-linear shock fronts resulting from the steepening of magneto-acoustic waves in sunspot umbrae \citep{2003A&A...403..277R,2006ApJ...640.1153C,2013A&A...556A.115D,2017ApJ...845..102H}. These shock fronts, commonly known as `umbral flashes' \citep[UFs;][]{1969SoPh....7..351B}, are ideal candidates to study subsequent chromospheric magnetic field fluctuations, since they are energetic, highly non-linear and display well defined properties, allowing their effects on the localized umbral magnetic field to be investigated. UFs exhibit a periodicity of approximately 3~minutes, which is a consequence of their source, in the form of upwardly propagating magneto-acoustic $p$-mode waves traversing the density stratification of the lower solar atmosphere and subsequently forming shocks \citep{1970SoPh...13..323H,2010ApJ...722..131F}. The shock fronts manifest observationally as blue-shifted emission due to the initially upward motion of the shocking plasma, followed by an observed red-shift resulting from the plasma returning to an equilibrium position as it radiatively cools \citep{1997ApJ...481..500C,2000SoPh..192..373B,2008A&A...479..213B,2009A&A...494..269V,2010ApJ...722..888B,2013A&A...553A..73B}. 

Advancements in observing techniques have revealed that UFs are not single bulk processes, but are instead a combination of small-scale interactions. High resolution Stokes profiles have shown that UF atmospheres are composed of two distinct components: hot, upwardly moving plasma superimposed on top of a quiescent background plasma \citep{2000Sci...288.1396S,2001ApJ...550.1102S,2005ApJ...635..670C,2014ApJ...786..137T}. This is consistent with recent simulations that assessed the potential energy output of shocks into the localized chromosphere \citep{2010ApJ...722..888B,2011ApJ...735...65F,2014ApJ...795....9F}. 

Comprehensive studies of the small-scale interactions in the solar atmosphere are dependent on the techniques used for inferring plasma properties from the incident radiation. Initial methods, which assumed local thermodynamic equilibrium (LTE) of the plasma, prohibited unequivocal accuracy in the assessment of plasma parameters as the atmosphere transitions from the dense photosphere into the more diffuse chromosphere \citep{1987ApJ...322..473S,1992ApJ...398..375R,2013A&A...549A..24B}. This led to the development of non-LTE methods that take into account the more intricate physics required to fully model chromospheric plasma. While such non-LTE methods provide the most accurately inferred parameters, they are unfortunately computationally intensive. This limitation can be seen in the work of \citet{2013A&A...556A.115D}, who were forced to restrict their study to only two frames of spatially degraded data to study UFs using the Non-LTE Inversion COde using the Lorien Engine \citep[NICOLE;][]{2015A&A...577A...7S} inversion routine because of the computational effort. For data acquired in the chromospheric {\HeI}~line, the HAnle and Zeeman Light code \citep[HAZEL;][]{2008ApJ...683..542A} provides a similar inversion tool that can be parallelized to run simultaneously across a number of processing cores, whilst employing the well-understood physics related to optical pumping, atomic level polarization, level crossings and repulsions, in addition to the intrinsic Zeeman, Paschen-Back and Hanle effects \citep{2007ApJ...655..642T, 2009ApJ...694.1364T, 2010ASSP...19..118T}. 

Here, we present the first large-scale statistical study of vector magnetic field perturbations, arising as a result of UFs in the chromospheric umbra of a sunspot, using high-resolution data products obtained with the Dunn Solar Telescope. Almost 100{\,}000 spectro-polarimetric {\HeI} profiles are analyzed using HAZEL to provide unique insights into the dynamic fluctuations of both the longitudinal and transverse components of the vector magnetic field.  

%%%%%%%%%%%%%%%%%%%%%%%%%%%%%%%%%%%%%%%
%%%%%%%%%%%%%%%%%%%%%%%%%%%%%%%%%%%%%%%
\begin{figure}
\centering
\includegraphics[width=\columnwidth, clip=true]{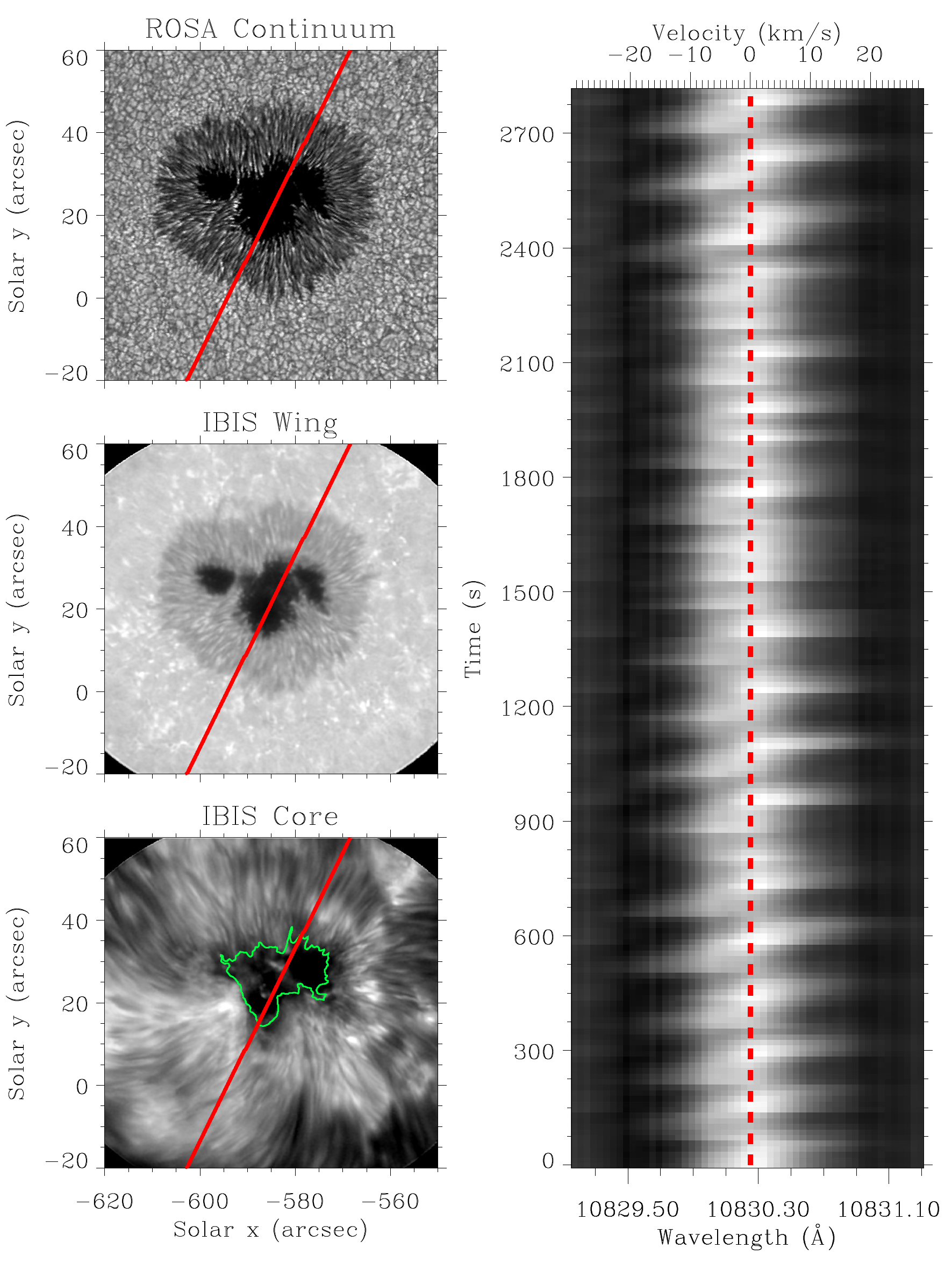}  
\caption{{\it{Top Left:}} ROSA 4170{\,}{\AA} continuum image of active region NOAA~12565. {\it{Middle Left:}} IBIS blue-wing snapshot acquired at 8540.82{\,}{\AA} (line core $-$ 1.3{\,}{\AA}). {\it{Bottom Left:}} IBIS {\CaIR} line-core image, where the green contour represents the location of the outer umbral boundary. In each panel, the solid red line represents the orientation and position of the FIRS spectral slit. {\it{Right Panel:}} Velocity--time image showing the spectral and temporal evolution of the {\HeI} Stokes $I$ line profile, where the black-to-white color scale represents the inverse spectral intensities to aid visual clarity. The vertical dashed red line represents the rest position of the {\HeI} line core.}
\label{fig:FOV}
\end{figure}
%%%%%%%%%%%%%%%%%%%%%%%%%%%%%%%%%%%%%%%
%%%%%%%%%%%%%%%%%%%%%%%%%%%%%%%%%%%%%%%

\newpage

\section{Observations and Data Reduction}{\label{sec:Observ}}
The data presented here represents an observational sequence obtained during 13:42 -- 14:30~UT on 2016 July 14 with the Dunn Solar Telescope (DST) at Sacramento Peak, New Mexico. The telescope was pointed towards active region NOAA~12565, positioned at heliocentric co-ordinates ($-$582{\arcsec},  30{\arcsec}), corresponding to a heliocentric angle of $38^\circ$ ($\mu \simeq$ 0.79), or N05.2E38.1 in the conventional heliographic co-ordinate system. Good seeing conditions were experienced throughout the observing period. Observations were obtained with three separate instruments: the Rapid Oscillations in the Solar Atmosphere \citep[ROSA;][]{2010SoPh..261..363J} imaging system, the Interferometric BIdimensional Spectrometer \citep[IBIS;][]{2006SoPh..236..415C} imaging spectrometer, and the Facility Infrared Spectropolarimeter \citep[FIRS;][]{2010MmSAI..81..763J} slit-based spectrograph. This study focuses primarily on the FIRS spectro-polarimetric and IBIS spectroscopic observations, with contextual 4170{\,}{\AA} continuum images provided by ROSA. 

The FIRS instrument was utilized to obtain diffraction-limited precision spectro-polarimetry of the {\HeI} line formed in the chromosphere. FIRS obtained {\HeI} spectra from a $75{\arcsec} \times 1{\,}.{\!\!}{\arcsec}125$ region of the solar disk, where the slit passed directly through the center of the sunspot umbra. A spatial sampling of $0{\,}.{\!\!}{\arcsec}15$ per pixel was obtained along the slit, while the width of the slit corresponded to $0{\,}.{\!\!}{\arcsec}225$. A 5-step raster of the umbral core was obtained by moving the slit $0{\,}.{\!\!}{\arcsec}225$ each step, producing a $1{\,}.{\!\!}{\arcsec}125$ wide slot. At each position, 12 consecutive modulation sequences were co-added to increase the signal-to-noise of each Stokes measurement, producing a total integration time of 14.6{\,}s per slit position. A total of 39 raster scans where performed over the observing period, equating to 195 individual slit steps. The spectral sampling for the {\HeI} line is $0.04${\,}{\AA}, and all resulting data was reduced and processed using the publicly available National Solar Observatory FIRS pipeline\footnote{For the FIRS reduction pipeline please visit \href{http://nsosp.nso.edu/dst-pipelines}{http://nsosp.nso.edu/dst-pipelines}.}. A slit-jaw camera, in-sync with the FIRS spectro-polarimetric exposures, was also employed to allow the precise spatial location and orientation of the FIRS slit to be mapped.

The IBIS instrument was employed to simultaneously sample the chromospheric Ca~{\sc{ii}} absorption profile at 8542.12{\,}{\AA}. IBIS employed a spatial sampling of $0{\,}.{\!\!}{\arcsec}098$ per pixel and imaged a circular field-of-view with a diameter of $97{\arcsec}$ centered on the leading sunspot of active region NOAA~12565. Forty-seven discrete, non-equidistant wavelength steps were used across the {\CaIR} line, covering the range 8540.82 -- 8543.42{\,}{\AA}, which resulted in a temporal cadence of 9.4{\,}s per imaging scan, with a total of 303 full spectral scans obtained. A whitelight camera, in sync with the narrow-band feed, was also employed to enable processing of the narrowband image sequence, alongside the implementation of high-order adaptive optics to improve quality \citep{2004SPIE.5490...34R}. Data reduction of the IBIS observations followed standard techniques (i.e., dark subtraction and flat fielding), yet also included a radial blue-shift correction that is required due to the use of a classical etalon mounting \citep{2008A&A...480..515C}, in addition to the alignment and de-stretching of the resulting image sequences \citep{2010ApJ...719L.134J, 2012ApJ...757..160J, Grant2018}. 

Simultaneous vector magnetograms were obtained from the Helioseismic and Magnetic Imager \citep[HMI;][]{2012SoPh..275..229S} on board the Solar Dynamics Observatory \citep[SDO;][]{2012SoPh..275....3P}. The outputs of the Very Fast Inversion of the Stokes Vector \citep[VFISV;][]{2011SoPh..273..267B} algorithm, applied to the HMI vector magnetogram data, were retrieved with a time cadence of 720{\,}s and a spatial sampling of $0{\,}.{\!\!}{\arcsec}5$. In addition, one contextual HMI 6173{\,}{\AA} continuum image, acquired at 13:41~UT, was obtained for the purpose of co-aligning the ROSA, IBIS and FIRS slit-jaw images with the HMI reference data. The vector magnetograms and continuum image incorporated corrections for scattered light, which has been documented by \citet{2016SoPh..291.1887C}, and further updated by \citet{2017arXiv170901593C}. Sub-fields of $200{\arcsec} \times 200{\arcsec}$ where extracted from the full-disk images, with a central pointing close to that of the ground-based observations. The HMI continuum image was then used to define absolute solar coordinates, with all ground-based observations subsequently subjected to cross-correlation techniques to provide sub-pixel co-alignment accuracy. The scattered light corrections made to the HMI data provided better visible fine-structuring of the umbral boundary and surrounding penumbral filaments, thus making the co-alignment with the ground-based data more accurate. 

Fully calibrated images obtained from the ROSA 4170{\,}{\AA} continuum, IBIS {\CaIR} blue-wing and IBIS {\CaIR} line-core datasets are displayed in Figure~{\ref{fig:FOV}}, alongside a corresponding time series of intensity spectra extracted from the FIRS {\HeI} observations.

\section{Analysis and Discussion}
\label{sec:Results}

\subsection{Flash Identification}
\label{sec:Flash}
The signatures of umbral flashes evolve rapidly in both wavelength and intensity. Depending on the formation height of the initial shock front, as well as the induced plasma properties, intensity variations will be observed across the corresponding spectral line profile. This is particularly evident in the right panel of Figure~{\ref{fig:FOV}}, where bright excursions can be seen extending far into the blue wing of the {\HeI} line. To better encapsulate the wide range of wavelengths that often display prominent UF-based brightenings, a number of wavelength-integrated IBIS images extending 0.2{\,}{\AA} into the blue-wing from the line core were created, thus establishing a pseudo-broadened filter width that better captures the dynamic spectral signatures of UFs in the {\CaIR} line. This ensures that the brightest part of each flash is included in the subsequent time series. 

Following the creation of wavelength-integrated images, the umbra was isolated from the penumbra and surrounding quiet Sun. The umbra and penumbra have to be segregated to ensure that no extraneous brightenings, such as penumbral jets \citep{2007Sci...318.1594K}, were included in the isolated time series. A time-averaged image of the wavelength-integrated dataset was created to provide a high contrast ratio between the umbra and surrounding penumbra. The resulting image was then manually thresholded to produce a contour of the umbral boundary. This was then turned into a binary mask, whereby all pixels within the umbra contour where assigned a value of `1', while all pixels outside the contour where assigned a value of `0'. The wavelength-integrated image sequence was then multiplied by the binary mask, leaving a purely umbral time series for subsequent study.

The identification of umbral flashes was carried out by applying a running mean subtraction method, similar to that employed by \citet{2003A&A...403..277R} and \citet{2015ApJ...800..129M}, to the wavelength-integrated time series defined above. Using a running mean allows long-duration time series to be more accurately normalized through the removal of brightenings that exist for longer time intervals than the UFs. The subtracted mean was calculated for each time step over the 15 images immediately preceding and following the image in question. This corresponded to an approximate $\pm$2.5 minute window. If the images occurred in the first or last 15 images of the dataset, they were subtracted by an average of the first or last 30 images, respectively.

%An advantage of using running mean subtraction is the removal of any small-scale intensity structuring that may have previously existed across the spatial extent of the sunspot.
With a normalized intensity, it is possible to use thresholding techniques to accurately detect and extract umbral flashes, which will have characteristic intensity excursions above the background. Following mean subtraction, the wavelength-integrated maps have an average background value of zero. Pixels corresponding to UFs were identified as intensity excursions exceeding 12$\sigma$ above the background, where $\sigma$ is the standard deviation of the umbral intensity time series. Such a large threshold ensures that the detected brightenings are statistically significant (i.e., not a consequence of detector noise or smaller-amplitude magneto-acoustic waves), with 298{\,}091 individual flash pixels identified in the IBIS time series across the $\sim$50-minute observational period. 

To map the detected UFs to the times and locations captured by FIRS, a co-registration process was implemented. Here, the IBIS and FIRS cadences of 9.4{\,}s and 14.6{\,}s, respectively, were used to match the IBIS UF detections to the corresponding FIRS spectra obtained closest in time. Next, the spatial location of the UFs detected in IBIS were mapped on to the corresponding FIRS slit-jaw image. If the mapped pixels lay outside of the FIRS slit they were excluded from subsequent study, while UFs that lay within the FIRS slit were noted. In total, the 298{\,}091 individual flash pixels detected in the IBIS dataset were reduced to 12{\,}988 individual spectra once mapped across to FIRS. Examination of the {\HeI} `sawtooth' profiles, identifiable in the right panel of Figure~{\ref{fig:FOV}}, revealed that the UFs detected in IBIS simultaneously impacted the corresponding FIRS spectra. This is not unexpected, since both spectral lines are chromospheric in nature, but this confirms the suitability of UF detections in IBIS as a marker for which spectro-polarimetric pixels to extract from the FIRS dataset for subsequent study. 

%%%%%%%%%%%%%%%%%%%%%%%%%%%%%%%%%%%%%%%
%%%%%%%%%%%%%%%%%%%%%%%%%%%%%%%%%%%%%%%
\begin{figure*}
\centering
\includegraphics[width=8.9cm, clip=true]{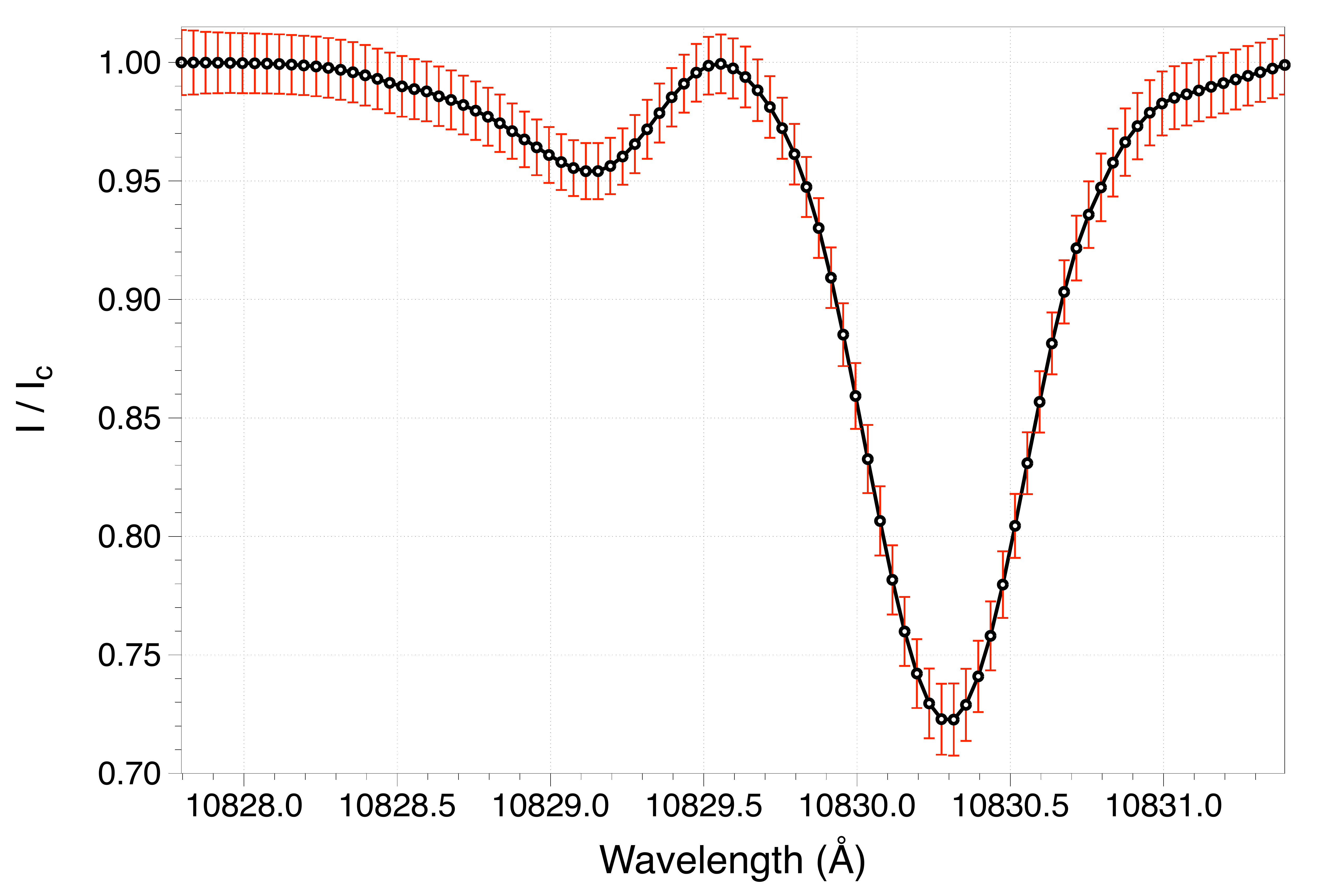}  
\includegraphics[width=8.9cm, clip=true]{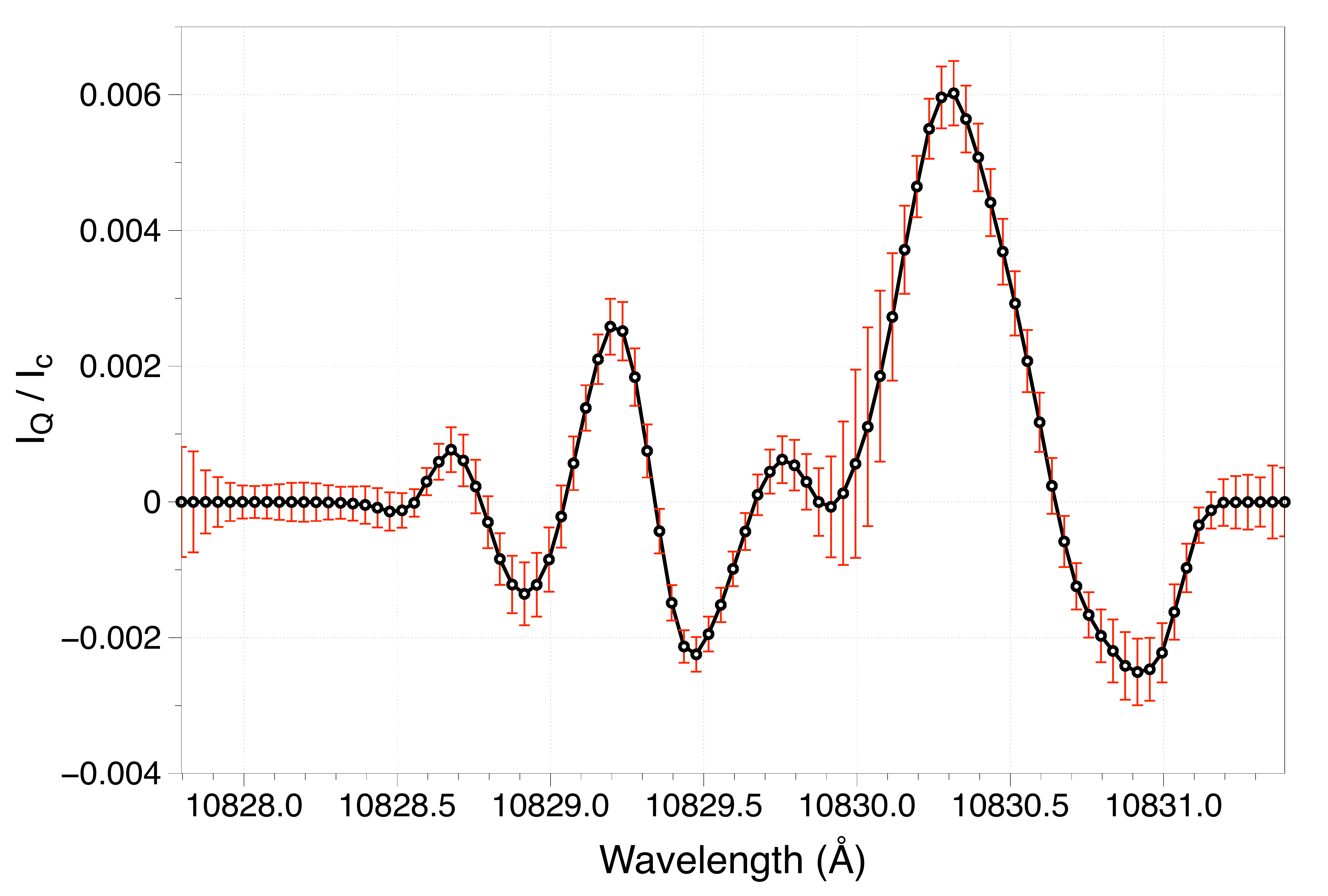} 
\includegraphics[width=8.9cm, clip=true]{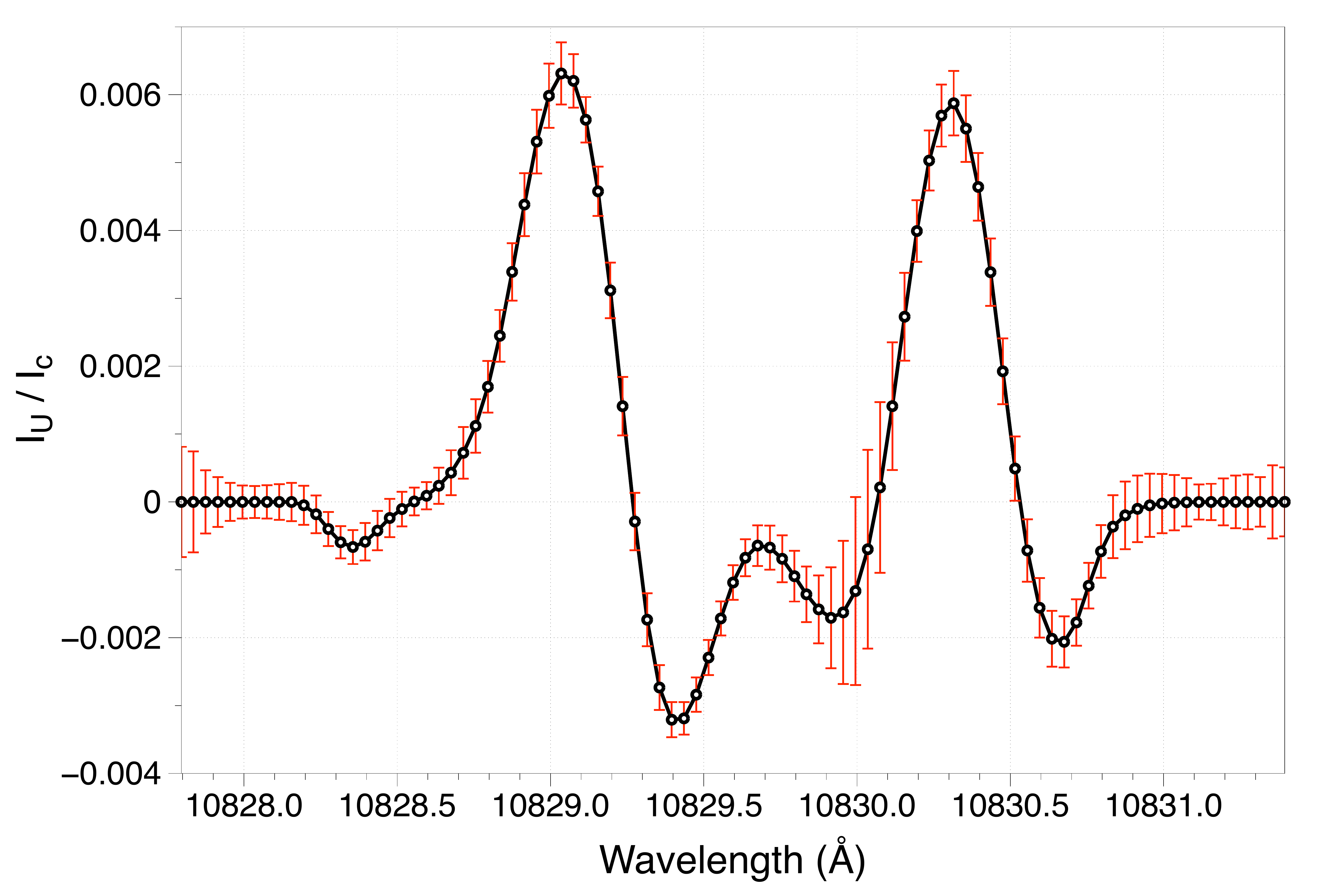} 
\includegraphics[width=8.9cm, clip=true]{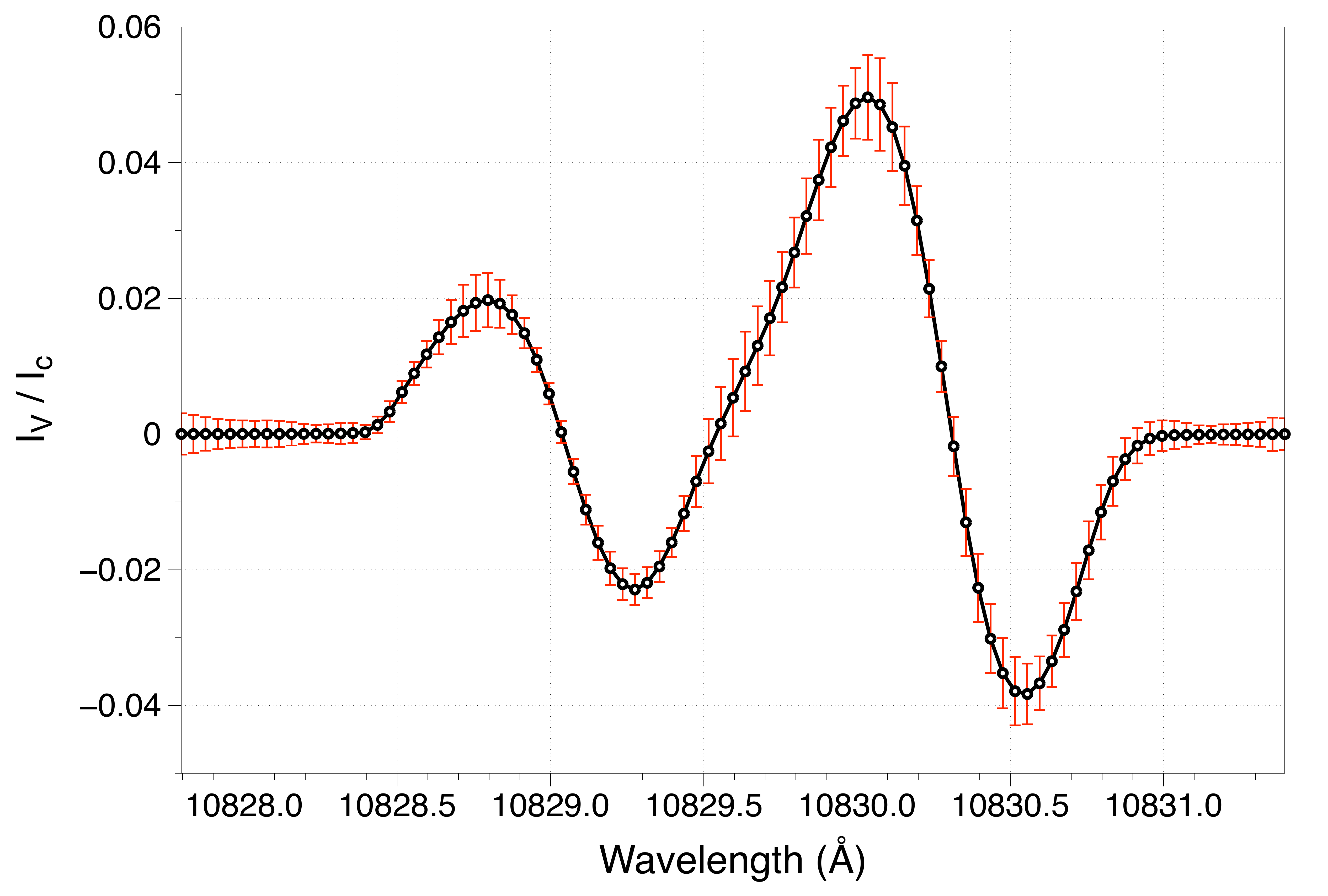} 
\caption{Clockwise from upper-left represents sample {\HeI} Stokes $I$, $Q$, $V$ and $U$ spectra (solid black lines), each normalized by the average continuum intensity found in the Stokes $I$ observations. Red error bars in each panel represent the spatially- and temporally-averaged standard deviations between the input FIRS and synthethized HAZEL intensities.}
\label{fig:Spectra}
\end{figure*}
%%%%%%%%%%%%%%%%%%%%%%%%%%%%%%%%%%%%%%%
%%%%%%%%%%%%%%%%%%%%%%%%%%%%%%%%%%%%%%%

%%%%%%%%%%%%%%%%%%%%%%%%%%%%%%%%%%%%%%%
%%%%%%%%%%%%%%%%%%%%%%%%%%%%%%%%%%%%%%%
\begin{figure*}
\centering
\includegraphics[width=8.9cm, clip=true]{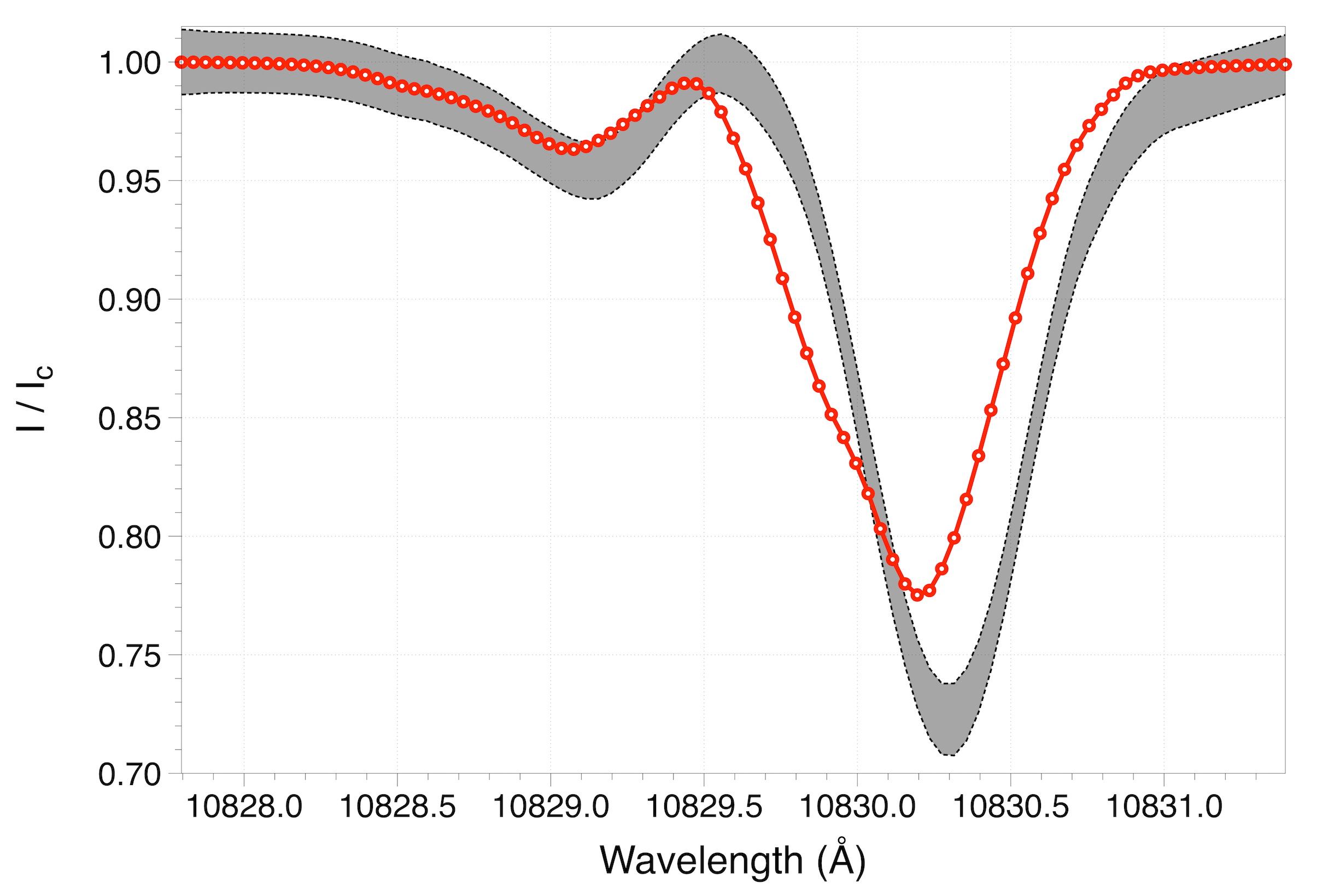}  
\includegraphics[width=8.9cm, clip=true]{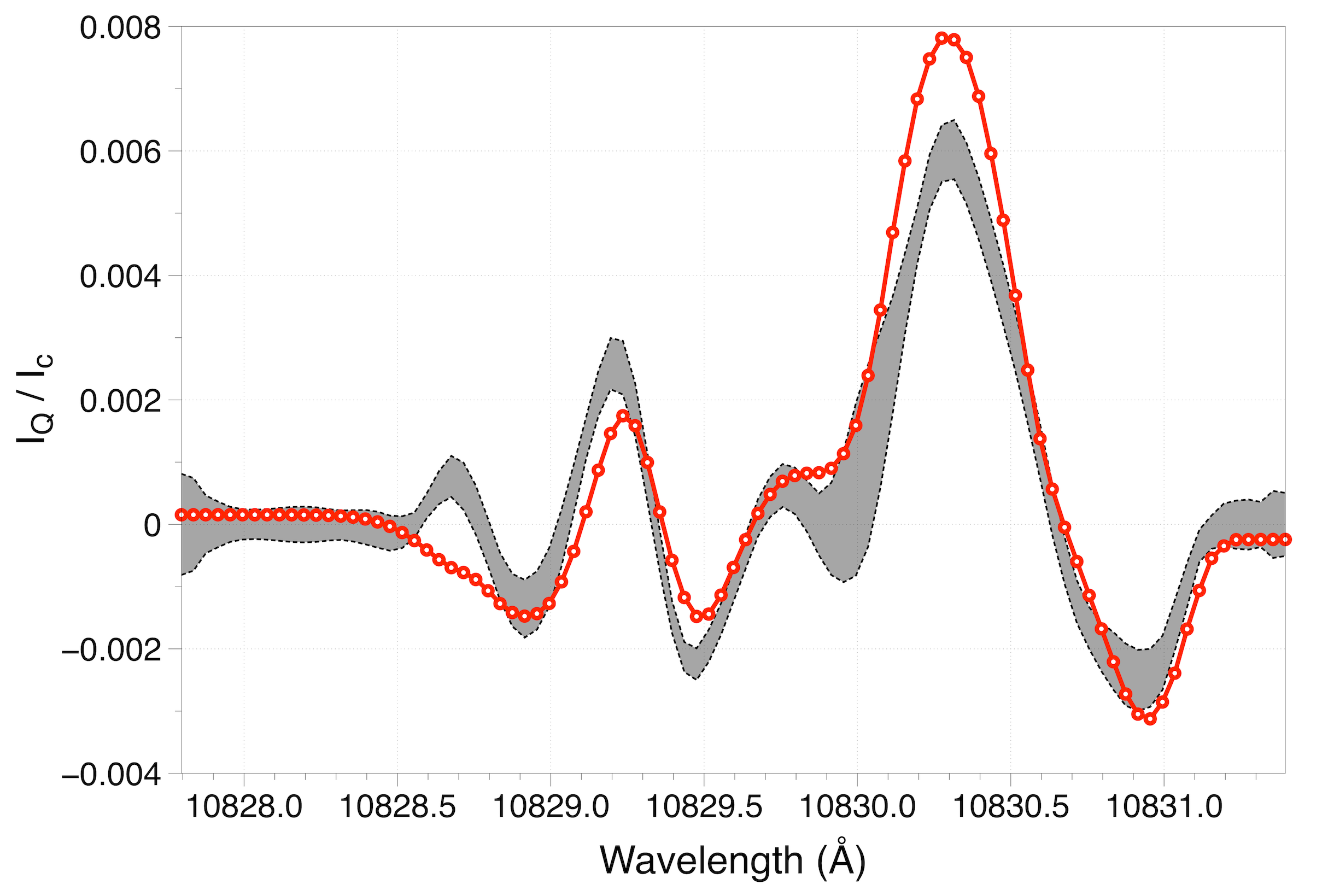} 
\includegraphics[width=8.9cm, clip=true]{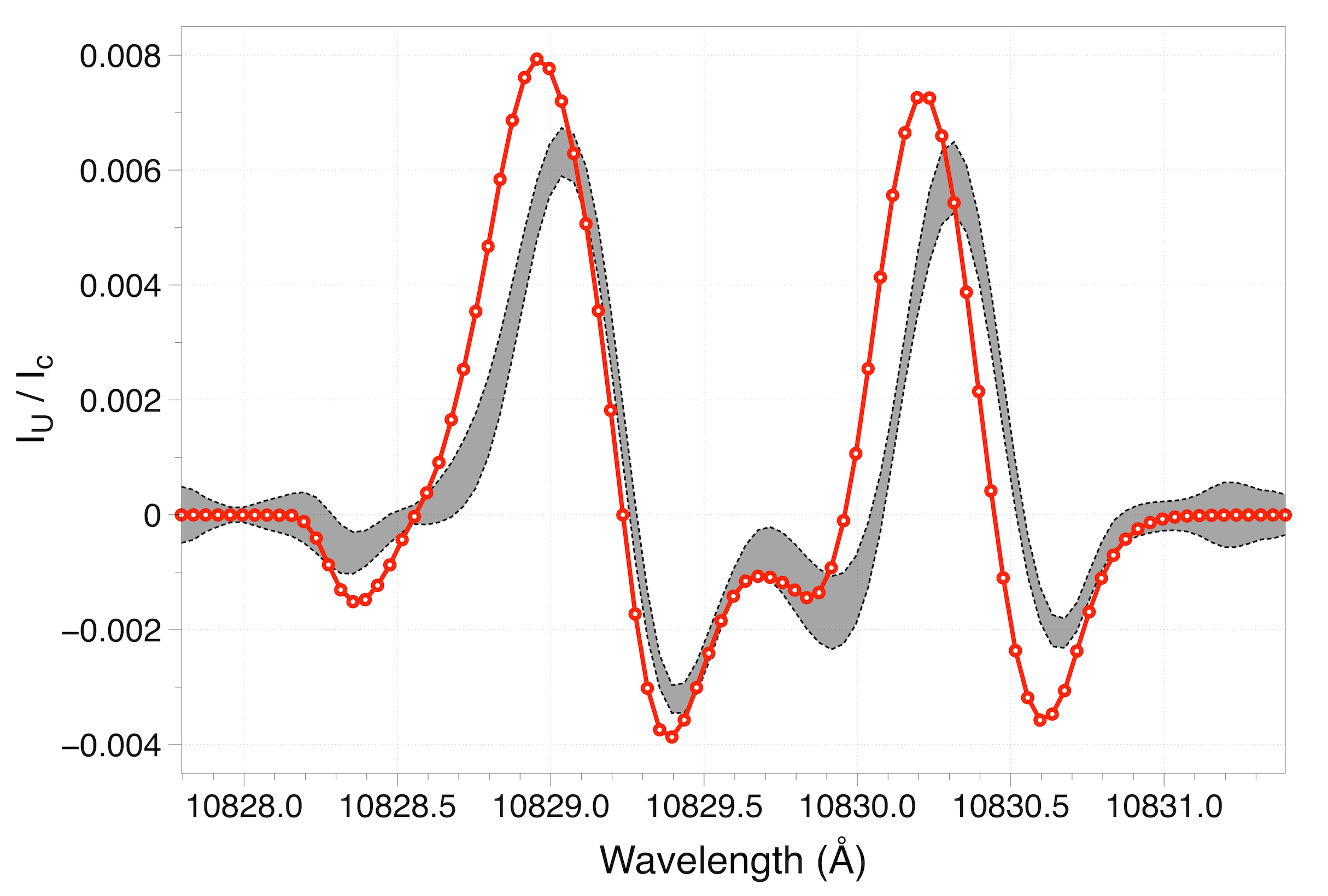} 
\includegraphics[width=8.9cm, clip=true]{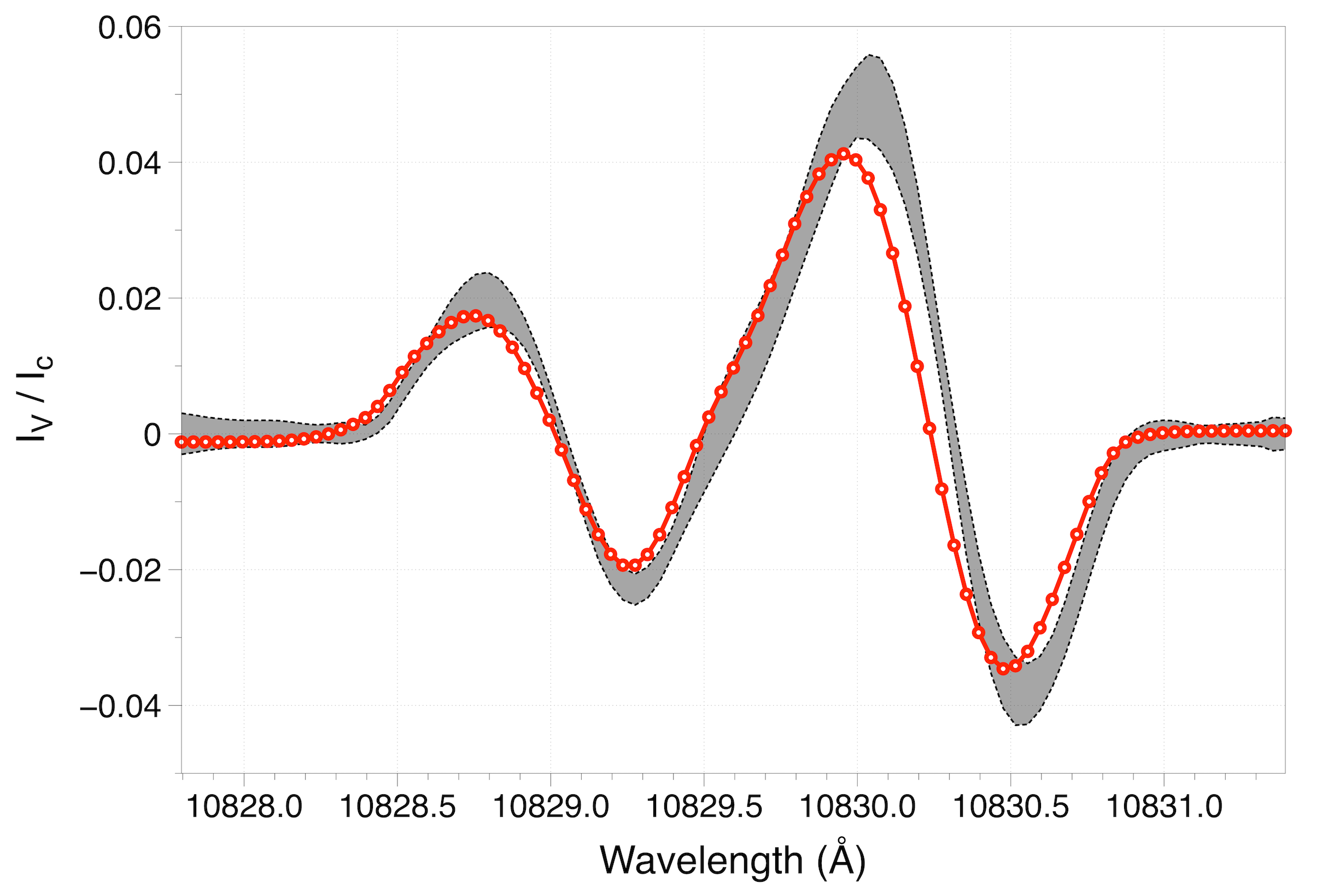} 
\caption{Clockwise from upper-left represents sample {\HeI} Stokes $I$, $Q$, $V$ and $U$ spectra corresponding to an UF (solid red lines), each normalized by the average continuum intensity found in the Stokes $I$ observations. Grey shaded regions in each panel represent the spatially- and temporally-averaged quiescent umbral profiles, including their respective 1$\sigma$ errors as depicted in Figure \ref{fig:Spectra}.}
\label{fig:Flash}
\end{figure*}
%%%%%%%%%%%%%%%%%%%%%%%%%%%%%%%%%%%%%%%
%%%%%%%%%%%%%%%%%%%%%%%%%%%%%%%%%%%%%%%

\subsection{HAZEL inversion code}
\label{sec:HAZEL}
To investigate the effects that UFs have on the local magnetic field, the co-spatial and co-temporal {\HeI} Stokes profiles extracted from FIRS were inverted using the HAZEL code. Following the FIRS data calibration, any residual fringes remaining in the spectra were removed following an approach based on the Relevance Vector Machine method \citep[RVM;][]{Tipping00therelevance}, similar to that used by \citet{2012A&A...547A.113A}. The observed spectra are decomposed as linear combinations in a non-orthogonal dictionary made up of sines and cosines of different frequencies (that are used to capture the fringes) and Gaussians of different widths and positions (that capture the spectral lines). The RVM computes linear combinations by imposing sparsity constraints on the coefficients. Given that the dictionaries used for spectral lines and fringes are largely incoherent, the sparsity constraints help avoid any mixing. The de-fringing is performed by subtracting the components of the linear combinations associated with sines and cosines, while keeping the rest. 

The efficiency of the parallelized HAZEL code allowed the entire FIRS dataset, consisting of 93{\,}991 individual {\HeI} Stokes spectra (12{\,}988 flashing pixels and 81{\,}003 quiescent spectra), to be inverted in approximately 36~hours using 10~CPU cores on a 2.4{\,}GHz Intel Xeon machine. HAZEL uses a forward modeling code with an efficient global optimization method for the inversion solution, and assumes a constant-property slab of He~{\sc{i}} atoms up to a height of 3{\arcsec} (or $\sim$2100~km) above the visible solar surface. It is assumed that all atoms within this slab are illuminated by unpolarized photospheric continuum radiation. This radiation subsequently produces population imbalances and quantum coherences between sub-levels. The atoms in the triplet system of {\HeI} are described by their first five terms, each one containing fine structure $J$-levels \citep{2008ApJ...683..542A}. The code solves the statistical equilibrium equations for a multi-term atom, in which quantum coherences are allowed among the different $J$-levels pertaining to the same term. A number of important physical parameters are output, notably the magnetic field strength, $B$, the inclination of the magnetic field vector, $\theta_B$, the azimuth of the vector magnetic field, $\chi_B$, the Doppler velocity of the embedded plasma, $v_{\mathrm{mac}}$, and the thermal (velocity) broadening of the sampled plasma, $v_{\mathrm{th}}$. 

The output for each of the spectral fits, alongside the parameters detailed above, is a synthetic profile, along with the corresponding fitment errors. Figures~{\ref{fig:Spectra}} and {\ref{fig:Flash}} document the quality of the synthetic HAZEL fits. Figure~{\ref{fig:Spectra}} displays the Stokes $I/I_{\mathrm{c}}$, $Q/I_{\mathrm{c}}$, $U/I_{\mathrm{c}}$ and $V/I_{\mathrm{c}}$ spectra corresponding to a non-flashing umbral pixel (solid black line), where I$_{\mathrm{c}}$ is the average continuum intensity found in Stokes~$I$. At each wavelength point, the red error bars indicate the spatially- and temporally-averaged standard deviations corresponding to the offsets between the real and synthetic intensities. The red lines in Figure~{\ref{fig:Flash}} display Stokes $I/I_{\mathrm{c}}$, $Q/I_{\mathrm{c}}$, $U/I_{\mathrm{c}}$ and $V/I_{\mathrm{c}}$ spectra corresponding to a shocking umbral pixel. The shaded grey regions represent the quiescent umbral spectra (alongside their associated 1$\sigma$ errors) from Figure~\ref{fig:Spectra}. The differences between quiescent and UF spectra are clearly apparent upon examination of Figure~\ref{fig:Flash}. Stokes $I/I_{\mathrm{c}}$ exhibits the characteristic blue-shifted and enhanced emission traditionally associated with UFs, resulting from the upward propagation and non-linear shocking of their host magneto-acoustic waves. Stokes $Q/I_{\mathrm{c}}$ and $U/I_{\mathrm{c}}$ profiles display larger amplitudes, highlighting that UFs are able to modulate the linear polarization signals associated with the {\HeI} spectral line. The amplitude of the Stokes $V/I_{\mathrm{c}}$ profile decreases as a consequence of the interplay between it and the Stokes $Q/I_{\mathrm{c}}$ and $U/I_{\mathrm{c}}$ spectral behavior. The lack of an observed polarity change in the Stokes $V/I_{\mathrm{c}}$ profile is indicative of the two-component atmosphere model for UFs, comprised of both quiescent and shocking plasma in the same spatial location \citep{2001ApJ...550.1102S}. A polarimetric uniformity across quiescent and UF Stokes profiles suggests that the less energetic phase of UF morphology is being sampled; a consequence resulting from the upper-chromospheric formation height of the {\HeI} spectral line \citep{1981ApJS...45..635V, 1994IAUS..154...35A}. This is in contrast to upper-photospheric and lower-chromospheric observations of UF phenomena, which are obtained close to the formation heights of the UFs themselves \citep{Grant2018}, hence producing a strong polarity change \citep{2013A&A...556A.115D, 2017ApJ...845..102H}. When considered on a statistical basis (i.e., not isolating individual profiles that may inadvertently bias subsequent analyses), the excellent quality of the FIRS data and synthetic HAZEL spectra means that a complete study of vector magnetic field fluctuations can be undertaken. 

Previous studies investigating magnetic field perturbations in the aftermath of UF phenomena have only focused on the line-of-sight (LOS) components of magnetic field due to inherently weak Stokes $Q$ and $U$ signals.  However, our accurately constrained Stokes $Q$ and $U$ profiles allow both the parallel ($B_{\mathrm{z}}$) and transverse ($B_{\mathrm{trans}}$) components of the magnetic field, with respect to the solar normal, to be mapped with a high degree of precision. In order to convert the $B$, $\theta_B$ and $\chi_B$ parameters into their parallel and transverse components, we adopt the methods documented by \citet{1990SoPh..126...21G}. Azimuthal disambiguation of the transverse magnetic field vectors was performed through comparison with the photospheric reference HMI vector magnetograms, and through use of the algorithms detailed by \citet{2011arXiv1104.1228R}. Furthermore, to study the thermal response of the umbra to the shocks, the thermal velocity broadening term was transformed into an absolute temperature, $T$, through the relation $T = v_{\mathrm{th}}^{2}M/2k$, where $M$ and $k$ are the atomic mass \citep{2005asto.book.....C} and the Boltzmann constant, respectively. It should be noted that the derived temperatures correspond to upper limits, since there may be unresolved microscopic motions, radiative transfer effects and unresolved turbulent velocities that are not taken into account following this assumption. 

\subsection{Umbral Flash Parameters}
\label{sec:UF}
%%%%%%%%%%%%%%%%%%%%%%%%%%%%%%%%%%%%%%%
%%%%%%%%%%%%%%%%%%%%%%%%%%%%%%%%%%%%%%%
\begin{figure}
\centering
\includegraphics[width=\columnwidth, clip=true]{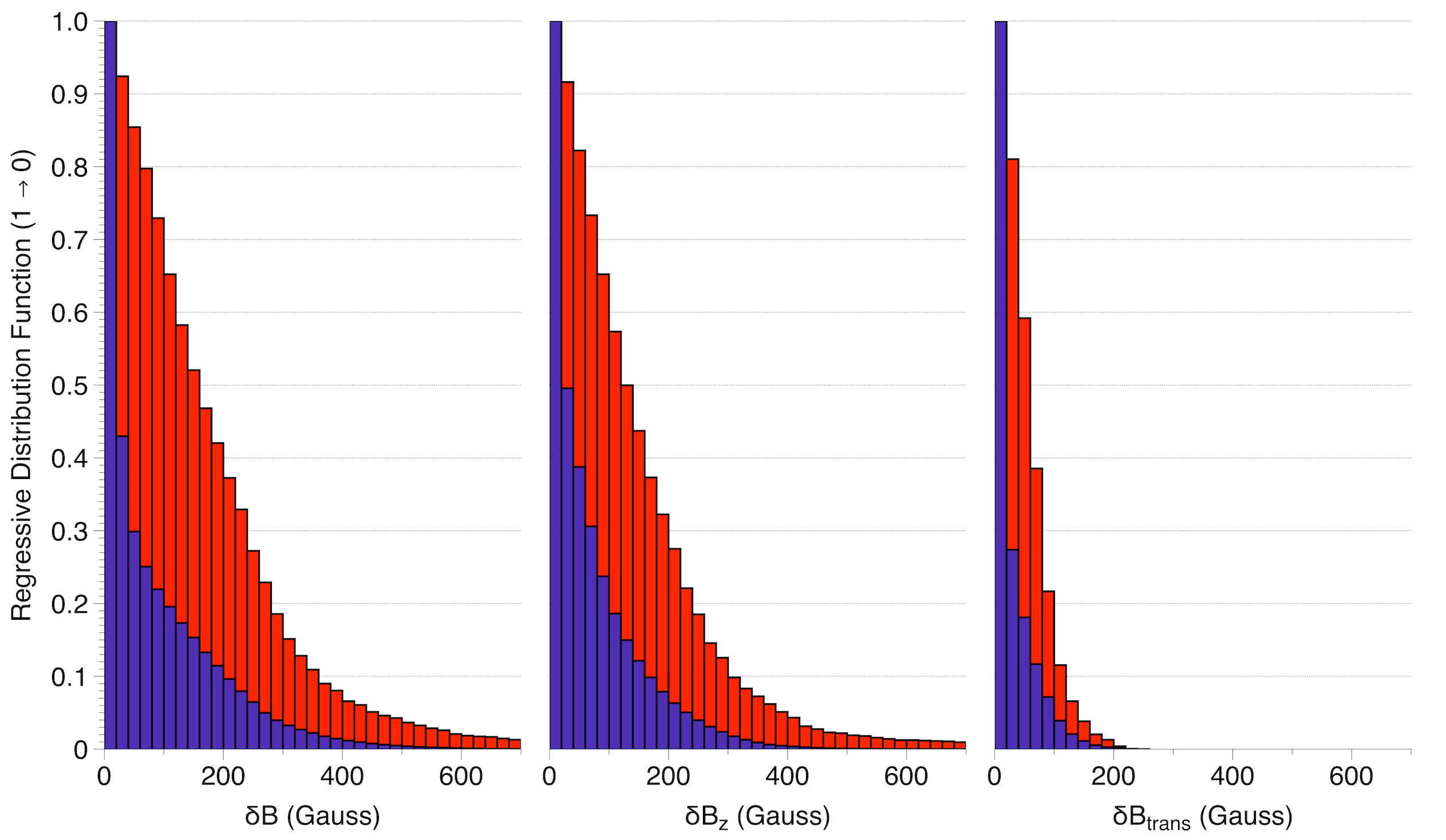} 
\includegraphics[width=\columnwidth, clip=true]{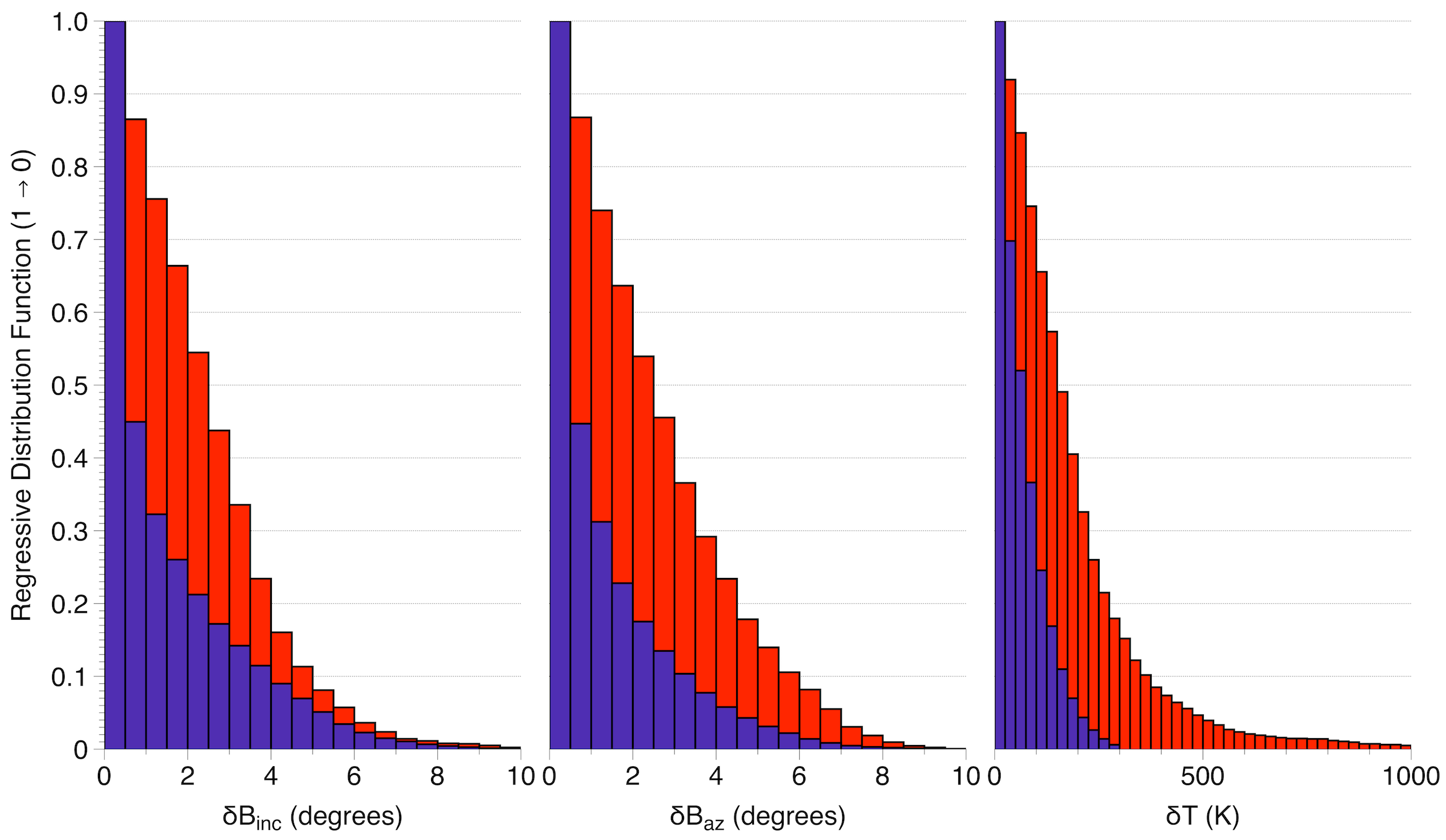} 
\caption{Regressive probability distributions comparing the change in measured plasma parameters between UF (red bars) and non-shocking (blue bars) umbral locations. Clockwise from upper-left are the absolute magnitude changes in the total magnetic field, the magnetic field component parallel to the solar normal, the field strength perpendicular to the solar normal, the inferred plasma temperature, and the azimuthal and inclination angles of the vector field, respectively.}
\label{fig:Histogram}
\end{figure}
%%%%%%%%%%%%%%%%%%%%%%%%%%%%%%%%%%%%%%%
%%%%%%%%%%%%%%%%%%%%%%%%%%%%%%%%%%%%%%%

Following the inversion of all FIRS spectro-polarimetric data, changes in various plasma parameters, with respect to their means for that pixel, were calculated for the entire dataset. Figure~{\ref{fig:Histogram}} details regressive probability distributions of the relevant parameters, notably the absolute magnitude changes in total magnetic field strength, $\delta{B}$, parallel field strength, $\delta{B}_{\mathrm{z}}$, transverse field strength, $\delta{B}_{\mathrm{trans}}$, field inclination angle, $\delta\theta_{B}$, field azimuth angle, $\delta\chi_{B}$, and the associated temperature, $\delta{T}$. In Figure~{\ref{fig:Histogram}}, the red bars relate to the pixels demonstrating UF phenomena, while the blue bars correspond to non-flashing umbral locations. The use of regressive histograms (i.e., displaying the probability, from $1 \rightarrow 0$, that the measured variable will take a value greater than or equal to the axis marker) allows the distribution shapes and morphologies between UF and non-UF locations to be much more easily compared, since the much higher number statistics associated with non-UF locations would dominate a standard occurrence histogram.
%%%%%%%%%%%%%%%%%%%%%%%%%%%%%%%%%%%%%%%
%%%%%%%%%%%%%%%%%%%%%%%%%%%%%%%%%%%%%%%
\begin{figure*}
\includegraphics[width=\textwidth, clip=true]{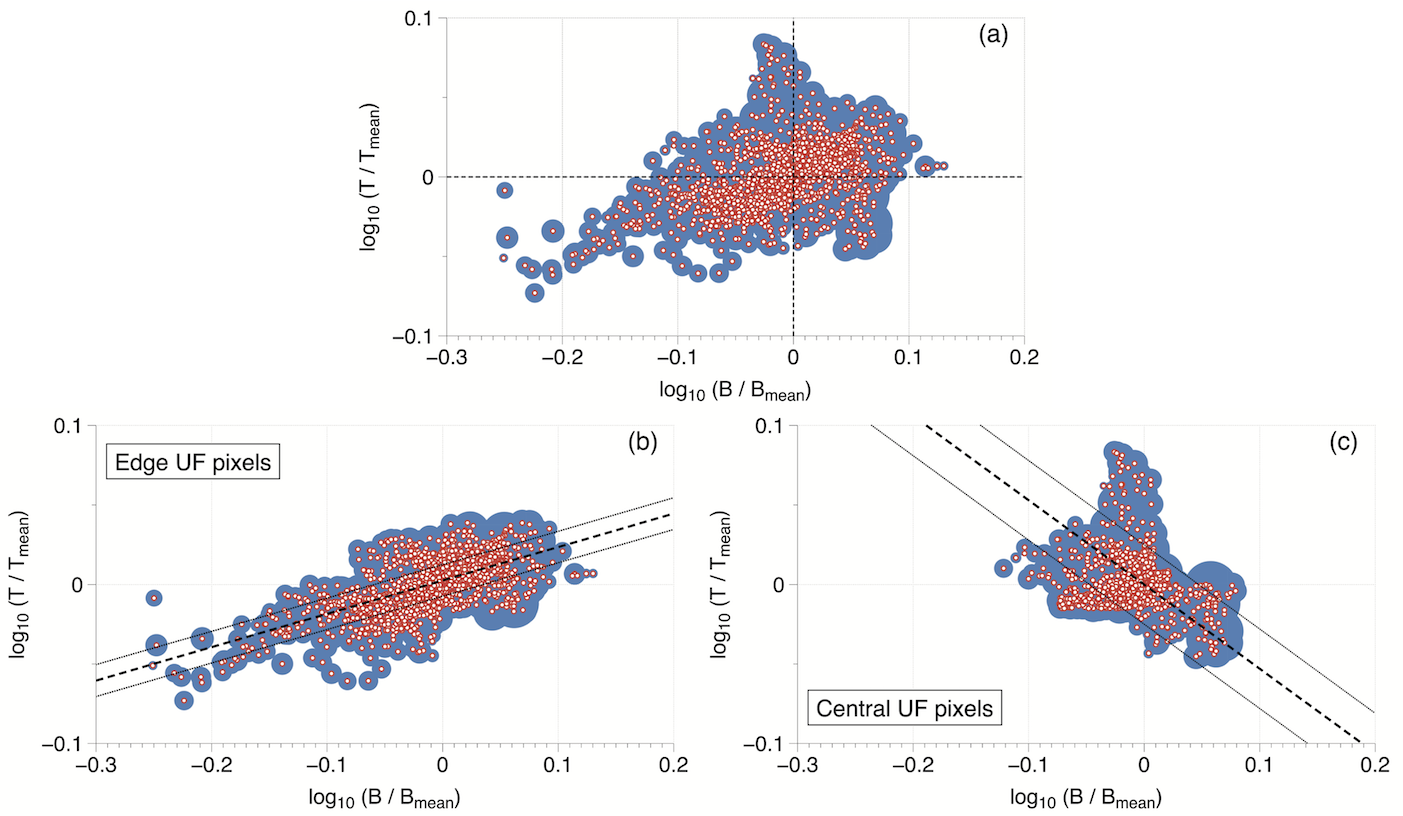}
\caption{Scatter diagrams of the total magnetic field strengths and temperatures induced by UFs, each normalized by their respective pixel means (red circles). The shaded blue regions represent the associated errors for each of the data points. Panel (a) represents all detected UFs, with the dashed black lines highlighting the origin location. Panel (b) displays only edge identified UF pixels, while panel (c) depicts centrally defined UF pixels. A linear line of best fit is plotted using a dashed black line, while the dotted black lines represent the $1\sigma$ error boundaries associated with the fitted line.}
\label{fig:linear}
\end{figure*}
%%%%%%%%%%%%%%%%%%%%%%%%%%%%%%%%%%%%%%%
%%%%%%%%%%%%%%%%%%%%%%%%%%%%%%%%%%%%%%%

Inspection of the upper-left and upper-middle panels of Figure~\ref{fig:Histogram} highlights that UFs cause noticeably larger perturbations in both the total magnetic field, $B$, and its vertical component, $B_{\mathrm{z}}$, when compared to more quiescent umbral locations, which are likely to be dominated by relatively small-amplitude, linear magneto-acoustic waves. This is in contrast to some of the early work of \citet{2013A&A...556A.115D}, who observed no evidence for magnetic field fluctuations resulting from UFs. When displayed on the same axis range (upper-right panel of Figure~{\ref{fig:Histogram}}), the fluctuations in the transverse component, $B_{\mathrm{trans}}$, of the magnetic field vector appear much smaller. This is likely a consequence of the initial shock front propagating along the wave vector of the underlying magneto-acoustic waves, which are directed along the magnetic field lines. Since the umbral magnetic fields are dominated by mostly vertical components (i.e., along $B_{\mathrm{z}}$), the initial strong perturbation of a developing shock will likely affect plasma along this motion path, hence producing $\delta{B}_{\mathrm{z}} > \delta{B}_{\mathrm{trans}}$, as can be seen in Figure~{\ref{fig:Histogram}}. However, importantly, the histograms associated with changes in the magnetic field (either vector or total) indicate that larger-amplitude deviations are found in UF locations when compared to their non-shocking atmospheric counterparts. 

The lower-left and lower-middle panels of Figure~\ref{fig:Histogram} display the absolute changes measured in the inclination and azimuthal direction of the magnetic field. It can be seen that for non-UF locations (blue bars), approximately 65\% of the inverted pixels display very small fluctuations from the background mean on the order of $\delta\theta_{B} \leq 1^{\circ}$ and $\delta\chi_{B} \leq 1^{\circ}$. This is in stark contrast to the UFs, where approximately 65\% of the inverted spectra demonstrate $\delta\theta_{B} \leq 3^{\circ}$ and $\delta\chi_{B} \leq 3^{\circ}$. This increase in magnetic field deflections, through a combination of inclination and/or azimuthal changes, highlights the impact developing shocks can have on their surrounding plasma. This is likely to be a consequence of the localized increase in adiabatic pressure resulting from the strongest UFs \citep{2017ApJ...845..102H}, which can subsequently deflect the surrounding magnetic field concentrations.

The lower-right panel of Figure~\ref{fig:Histogram} highlights the large temperature excursions resulting from UFs occurring in the relatively cool surrounding umbral plasma. UF locations induce temperature excursions spanning a few tens of degrees, up to $\sim$$1100$~K, while non-UF pixels demonstrate a much smaller range of fluctuations extending from $\sim$$0 - 300$~K. The absolute range of UF temperature fluctuations is in agreement with the work of \citet{2013A&A...556A.115D}, who observed temperature increases of $\sim$$1000$~K. However, this is more towards the upper end of our temperature fluctuation distribution, which has a mean value on the order of a few hundred~K. This may be a consequence of the different spectral lines used, with \citet{2013A&A...556A.115D} performing their inversions on spectro-polarimetric {\CaIR} observations. Subtle variations in the formation heights between the {\HeI} and {\CaIR} spectra \citep{2010ApJ...722..131F} may directly influence the localized heating potential of the UFs. Furthermore, the {\HeI} plasma is often treated in an optically thin manner, and as a consequence, is less responsive to temperature fluctuations when compared to the optically thick lower solar atmosphere \citep{1997ApJ...489..375A}. \citet{Grant2018} showed that the strongest UFs can form as low as $\sim$$250$~km above the solar surface. Therefore, at the formation height of {\HeI} \citep[expected to be up to $\sim$$2100$~km;][]{1981ApJS...45..635V, 1994IAUS..154...35A}, the induced shock signatures will have abated, thus providing less energy dissipation, in the form of heat, to the upper chromosphere. This remains consistent with the hypothesis that the relatively high formation height of the {\HeI} spectral line (compared to other chromospheric absorption lines, e.g., {\CaIR}) naturally provides a two-component UF atmosphere, thus minimizing polarity changes in the observed spectro-polarimetric signals. 

\subsection{Relating Temperature and Magnetism}
\label{sec:Linear}
As can be seen from Figure~{\ref{fig:Histogram}}, UFs provide bigger induced magnetic field and temperature fluctuations  when compared to quiescent background umbral locations. However, Figure~{\ref{fig:Histogram}} plots the absolute changes for each parameter in order to form a comprehensive statistical picture. Therefore, in order to probe the relationship, if any, between the shock-induced magnetic field and temperature perturbations, it becomes necessary to map their direct, un-signed characteristics. The upper panel of Figure~{\ref{fig:linear}} displays a scatter diagram corresponding to the un-signed temperature and total magnetic field strength fluctuations for UF locations, each normalized by their respective pixel means. The blue shaded regions represent the associated errors provided by the HAZEL code for each data point, which are obtained following the numerical recipes documented by \citet{Press:1992:NRC:148286}. The covariance matrix is computed once a solution is found, with the errors subsequently derived. The fundamental assumption is that the $\chi^2$ surface is well reproduced as a multi-dimensional ellipsoid, or in other words, that the likelihood function is Gaussian with respect to the selected covariance matrix. An initial examination of this scatter diagram suggests the presence of two distinct populations: (1) a strong linear correlation between temperature and magnetic field fluctuations, and (2) a trend implying temperature increases are related to decreases in the local magnetic field strength (i.e., an anti-correlation). 

To investigate how the locations of the UFs, as positioned on the FIRS spectral slit and subsequently inverted using the HAZEL code, affect the scatter diagram, we isolate two distinct varieties of UF pixels: (1) those identified as `central' UF pixels, whereby they are bounded by positive UF pixel identifications on each side, and (2) those quantified as `edge' UF pixels, which demonstrate a non-flashing pixel in one (or both) neighboring pixels. By plotting just the edge pixels (Figure~{\ref{fig:linear}}b) and the central pixels (Figure~{\ref{fig:linear}}c), it becomes clear that the two distinct populations present in the original scatter diagram are governed by their characterization as either `edge' or `central' pixels. For each of the lower panels in Figure~{\ref{fig:linear}} a line of best fit is displayed using a dashed black line, while the dotted black lines indicate the $1\sigma$ uncertainties associated with the least-squares fitted line. These lines of best fit highlight the correlation and anti-correlation between the temperature and magnetic field fluctuations for `edge' and `central' UF pixel identifications, respectively. Due to the increased scatter of the pixels identified as central UF components, the $1\sigma$ uncertainties associated with the line of best fit are naturally larger. As one would expect, each line of best fit passes through the origin. 

\vspace{3mm}
\subsubsection{The Positively-correlated Temperature and Magnetism Relationship}
\begin{figure*}
\includegraphics[width=\textwidth, clip=true]{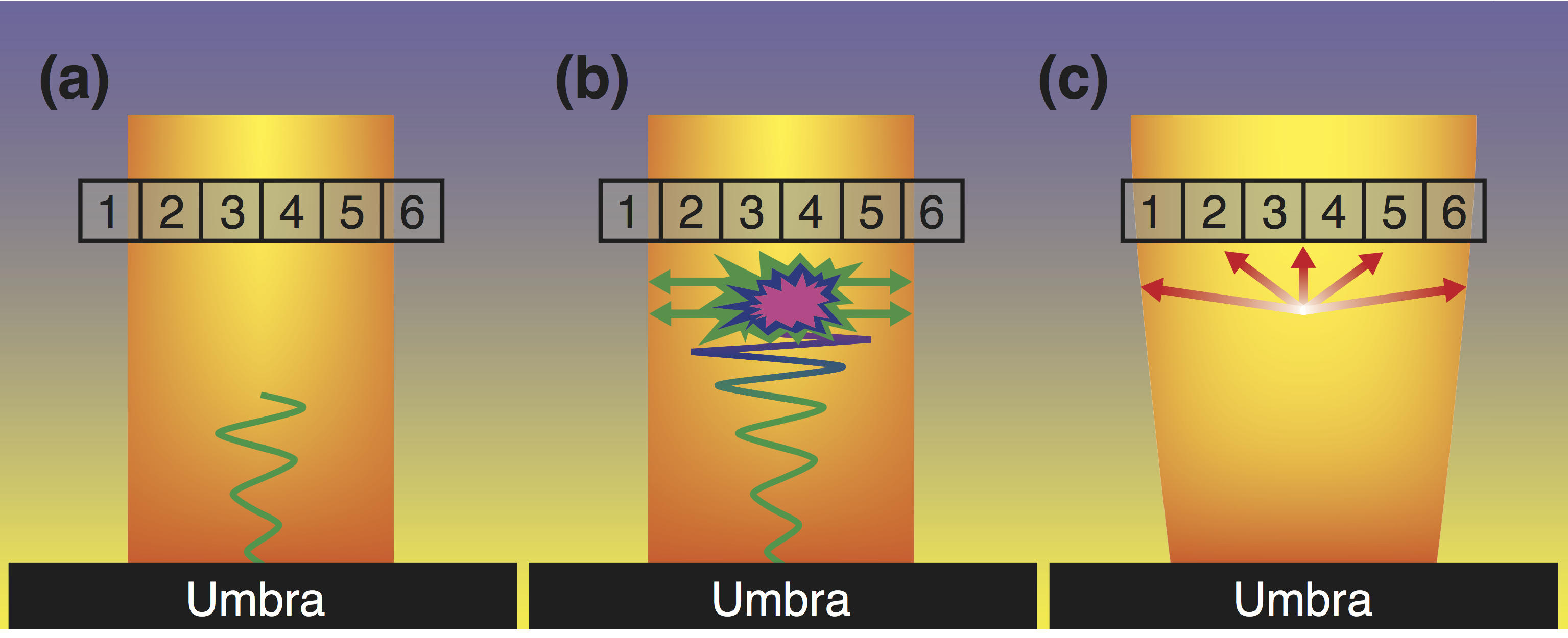}
\caption{A cartoon schematic depicting the physics responsible for the trends displayed in Figure~{\ref{fig:linear}}, whereby temperature enhancements for UF perimeter pixels correlate with magnetic field strength increases, while temperature enhancements for central UF pixels demonstrate reductions in the magnitude of the local magnetic field. Panel (a) represents the initial structuring of a magnetic flux tube embedded within the sunspot umbra, which channels the upward propagation of magneto-acoustic waves (green line). The translucent grey squares, labelled 1--6, represent six pixels positioned across the diameter of the magnetic flux tube. Panel (b) depicts the the steepening of the upwardly propagating magneto-acoustic waves, which ultimately develop into non-linear shock phenomena, producing increased adiabatic pressure acting outwards on the magnetic flux tube (green arrows). The increased adiabatic pressure causes the magnetic flux tube to expand (panel c) while the localized temperature is also increased as a result of the shock dissipation. Importantly, the magnetic field expands into the UF edge pixels (pixels 1 and 6), resulting in a correlation between the temperature and the magnetic field strength (lower-left panel of Figure~{\ref{fig:linear}}), while the central pixels (pixels 2--5) experience elevated temperatures alongside a net magnetic flux decrease due to the overall expansion of the magnetic flux tube (lower-right panel of Figure~{\ref{fig:linear}}).}
\label{fig:cartoon}
\end{figure*}

From Figure~{\ref{fig:linear}}b, it is clear that the temperature and magnetic field fluctuations identified at the edge of an UF event are closely correlated with one another. This effect can be readily visualized through the schematic displayed in Figure~{\ref{fig:cartoon}}. Here, a magnetic flux tube is anchored in the sunspot umbral core, which guides upwardly propagating magneto-acoustic waves. The embedded flux tube scenario is consistent with the observations, models and schematics put forward by \citet{1959SvA.....3..214S}, \citet{1979ApJ...234..333P}, \citet{1999A&A...347L..27S} and \citet{2002Natur.420..390T}, to name but a few. In Figure~{\ref{fig:cartoon}}, the translucent grey boxes labelled 1--6 highlight six example FIRS slit pixels across the diameter of the magnetic flux tube. Under initial conditions, pixels 2--5 demonstrate 100\% filling factors in relation to the observed flux tube (i.e., would be identified as `central' pixels), while pixels 1 and 6 are only fractionally filled by the magnetic flux tube (i.e., would be characterized as `edge' pixels). Then, the upwardly propagating magneto-acoustic waves begin to steepen, ultimately forming a non-linear shock in the form of an umbral flash. As depicted in Figure~{\ref{fig:cartoon}}b, this non-linear event provides increased adiabatic pressure. This pushes outwards on the walls of the magnetic flux tube, causing it to expand, at the same time as dissipation of the shock front induces increases in the localized plasma temperatures. Of particular note, as revealed in Figure~{\ref{fig:cartoon}}c, pixels 1 and 6 will experience both increases in the computed plasma temperatures (due to the dissipation of the UF shock front) and increases in the strength of the local magnetic field, resulting from the magnetic field of the expanding flux tube superimposing on top of the background umbral field. 

%%%%%%%%%%%%%%%%%%%%%%%%%%%%%%%%%%%%%%%
%%%%%%%%%%%%%%%%%%%%%%%%%%%%%%%%%%%%%%%
%%%%%%%%%%%%%%%%%%%%%%%%%%%%%%%%%%%%%%%
%%%%%%%%%%%%%%%%%%%%%%%%%%%%%%%%%%%%%%%

Such a linear relationship can be explained through thermodynamic considerations of the edge UF pixels. From examination of Figure~{\ref{fig:cartoon}}, the shock-induced increase in adiabatic pressure causes the magnetic flux tube to expand into the surrounding background plasma. For locations defined as `edge' UF detections, this produces temperature fluctuations, with the magnetic pressure allowed to vary through the increased magnetic field strength in that location (Figure~{\ref{fig:cartoon}}c), producing detectable perturbations in the derived strength of the magnetic field. This, of course, assumes that the umbral atmosphere is dominated by magnetic pressure (i.e., the plasma-$\beta \ll 1$), which is consistent with previous observational studies \citep[e.g.,][]{2013ApJ...779..168J, 2016ApJ...826...61A, Grant2018}. According to thermodynamic theory for adiabatic expansions, pressure fluctuations produce a subsequent change in temperature through the relationship,
\begin{equation}
\label{adabatic_eqn}
T_{2} = T_{\mathrm{mean}}{\Bigg(}\frac{P_{2}}{P_{\mathrm{mean}}}{\Bigg)}^{\frac{\gamma-1}{\gamma}}  \ ,	
\end{equation}
where $\gamma$ is the adiabatic index, $P_{\mathrm{mean}}$ and $T_{\mathrm{mean}}$ are the initial plasma pressure and temperature, respectively, of the average background plasma, while $P_{2}$ and $T_{2}$ represent the perturbed plasma pressure and temperature, respectively, resulting from the adiabatic process. Within the magnetically dominated environment of the sunspot umbra, the plasma will be dominated by magnetic pressure, $P_{M}$, defined as,
\begin{equation}
P_{M} = \frac{B^{2}}{2\mu_{0}} \ ,
\end{equation}
where $B$ is the total magnetic field strength and $\mu_{0}$ is the magnetic permeability. This allows Equation~{\ref{adabatic_eqn}} to be rewritten as,
\begin{equation}
\label{B_equation}
T_{2} = T_{\mathrm{mean}}{\Bigg(}\frac{B_{2}}{B_{\mathrm{mean}}}{\Bigg)}^{2\frac{\gamma-1}{\gamma}}  \ ,		
\end{equation}
where $B_{\mathrm{mean}}$ and $B_{2}$ are the initial average background and modified magnetic field strengths, respectively. Since the modified field strength is the result of a perturbation on top of the pre-existing background field (e.g., an increase in the local magnetic field strength caused by the adiabatic expansion of the magnetic flux tube as it crosses into the edge pixels, or vice versa), Equation~{\ref{B_equation}} can be subsequently rewritten by taking the logarithm of both sides as,
\begin{equation}
\label{B_equation_simple}
\log_{10}{\Bigg(}\frac{T_{2}}{T_{\mathrm{mean}}}{\Bigg)} = 2{\frac{\gamma-1}{\gamma}} \log_{10}{\Bigg(}\frac{B_{2}}{B_{\mathrm{mean}}}{\Bigg)} \ .
\end{equation}

By plotting the temperature and magnetic field perturbations calculated by the HAZEL code and defined in Equation~{\ref{B_equation_simple}}, Figure~{\ref{fig:linear}}b reveals a very clear linear relationship. The gradient of the line of best fit is calculated to be $2(\gamma-1)/\gamma = 0.22\pm0.01$, providing a value for the adiabatic index of $\gamma = 1.12\pm0.01$. \citet{2011ApJ...727L..32V} employed coronal EUV observations from the Extreme-ultraviolet Imaging Spectrometer \citep[EIS;][]{2007SoPh..243...19C} on board Hinode \citep[][]{2007SoPh..243....3K} to investigate the interplay between density and temperature perturbations found in magneto-acoustic slow-mode waves. \citet{2011ApJ...727L..32V} calculated the adiabatic index to be $\gamma = 1.10\pm0.02$, which is very close to the value determined here for chromospheric umbral locations. While the EUV coronal observations documented by \citet{2011ApJ...727L..32V} may be orders-of-magnitude hotter than those presented here, there are a number of similarities, namely (1) both environments will be dominated by magnetic pressure (i.e., plasma-$\beta < 1$), (2) as UFs begin to occur close to the atmospheric temperature minimum, both locations are likely to have significant temperature gradients that support thermal conduction, and (3) the outer `edge' pixels of an UF are likely to be less non-linear than those towards the center of the shocking region, hence remaining comparable to the linear wave modes examined by \citet{2011ApJ...727L..32V}. Furthermore, the numerical models of \citet{2015A&A...580A.110V} demonstrate that typical umbral densities ($\sim 10^{-8}$~g/cm$^{-3}$) and temperatures ($\sim 10^{4}$~K) are consistent with an adiabatic index of $1.1 \lesssim \gamma \lesssim 1.2$. Indeed, as documented by \citet{2004ApJ...616.1232K}, the relatively short period of the magneto-acoustic waves driving UFs may also stipulate an approximately isothermal (i.e., $\gamma \sim 1$) atmosphere, which lends credence to our derived value for the adiabatic index of $\gamma = 1.12\pm0.01$. 

From the gradient of the slope present in Figure~{\ref{fig:linear}}b, it can be seen that relatively small fluctuations in temperature cause large fluctuations in the magnetic field. Previous chromospheric studies that examined temperature fluctuations related to UFs found that brighter, more intense shocks resulted in greater temperature increases above the quiescent umbra. In addition to the adiabatic processes outlined above, an additional explanation for this is that bulk up-flows from the shocking plasma induce changes in the local plasma density, and thus the opacity of the shock-forming region \citep{2001ApJ...550.1102S}. As HAZEL derives a single optical depth value for each profile, through the integration of line-of-sight opacities, we are unable to verify or refute the role localized opacities play in our observed temperature excursions above the mean. However, as previously discussed, the typically high formation height of the {\HeI} spectral line \citep[$\sim$$2100$~km;][]{1981ApJS...45..635V, 1994IAUS..154...35A} is likely to result in the observed Stokes profiles capturing the less energetic phases of umbral flash morphology. As a result, the ensuing density fluctuations captured by {\HeI} spectra may be relatively minor, hence minimizing the amount of induced opacity fluctuations along the given line-of-sight. Nevertheless, under ideal conditions, the plasma emission should be hottest at the point of shock formation, with subsequent plasma cooling established as it decouples from the magnetic field and is allowed to move isotropically through the umbral atmosphere. 

\subsubsection{The Anti-correlated Temperature and Magnetism Relationship}
Figure~{\ref{fig:linear}}c, again, displays a predominantly linear relationship for central UF identifications, whereby now decreases in the total magnetic field strength correlate with relatively large increases in the corresponding plasma temperature, which is in agreement with the results put forward by \citet{2017ApJ...845..102H}. These particular findings are consistent with the theoretical viewpoint of shock formation in the lower solar atmosphere. Following the nomenclature of \citet{1997ApJ...481..500C}, a formal description of the emergent intensity, $I_{\nu}$, from a column height extracted from the umbral atmosphere (whose lower and upper boundaries are denoted by $z_{0}$ and $z_{1}$, respectively) is given by,
\begin{equation*}
I_{\nu} = \int_{z_{0}}^{z_{1}} S_{\nu}{\,}e^{-\tau_{\nu}}{\,}\chi_{\nu}{\,}dz \ ,
\end{equation*}
where $S_{\nu}$ is the source function, $e^{-\tau_{\nu}}$ is an exponential attenuation factor and $\chi_{\nu}{\,}dz$ is the product of the cross section and the column density of the emitters. Since the HAZEL inversions operate within a fixed range of optical depths, the exponential attenuation factor will remain constant throughout the evolution of each shock event. Therefore, the two factors capable of modifying the intensity are the source function and the column parameters of the emitters. Though small variations occur in the column densities at equivalent optical depths due to, e.g., the Wilson depression \citep[e.g.,][]{2013A&A...558A.130S}, its effect on the emergent intensity will be negligible when compared to the source function. \citet{1997ApJ...481..500C} showed that the source function varies as a function of atmospheric height, being largest in lower, photospheric locations where it is naturally coupled to the Planck function, and decreasing upwards into the chromosphere. Therefore, a consequence of this will be that the most intense UFs occur deepest in the umbral atmosphere. As a result, the deep-forming shock fronts have the ability to more significantly perturb the overlaying magnetic field lines. Here, the adiabatic plasma pressure of the more energetic upwardly propagating shocks pushes the magnetic field lines more strongly apart, thus reducing the magnitude of the measured magnetic field strength when observed in the `central' pixels at the formation height of the {\HeI} line. 

This can be visualized in Figure~{\ref{fig:cartoon}}, where in the aftermath of an UF event the original magnetic field flux is distributed across a larger area (e.g., by expanding into pixels 1 and 6, and potentially beyond). Thus, for pixels 2--5 in Figure~{\ref{fig:cartoon}}c, the increases in localized plasma temperatures will be correlated with decreases in the local magnetic field flux, hence producing an anti-correlation between the plasma temperature and the embedded magnetic field. The magnetic flux tube filling factors for pixels 2--5 remain at 100\% throughout the evolution of an UF event, hence the volume filling of the magnetic flux tube as a consequence of the increased adiabatic pressure produces a decrease in the localized magnetic field flux. Then, as the shock front begins to dissipate, the gravitational infall of the cooling plasma creates a pressure `vacuum', causing magnetic field lines to condense through a process similar to convective collapse in the quiet Sun. Therefore, UFs can be considered as a mechanism that causes amplification of the magnetic field perturbations displayed at a much weaker level by linear magneto-acoustic wave interactions with the chromospheric umbral plasma.  

%%%%%%%%%%%%%%%%%%%%%%%%%%%%%%%%%%%%%%%
%%%%%%%%%%%%%%%%%%%%%%%%%%%%%%%%%%%%%%%
\begin{figure*}
\centering
\includegraphics[width=0.95\columnwidth, clip=true]{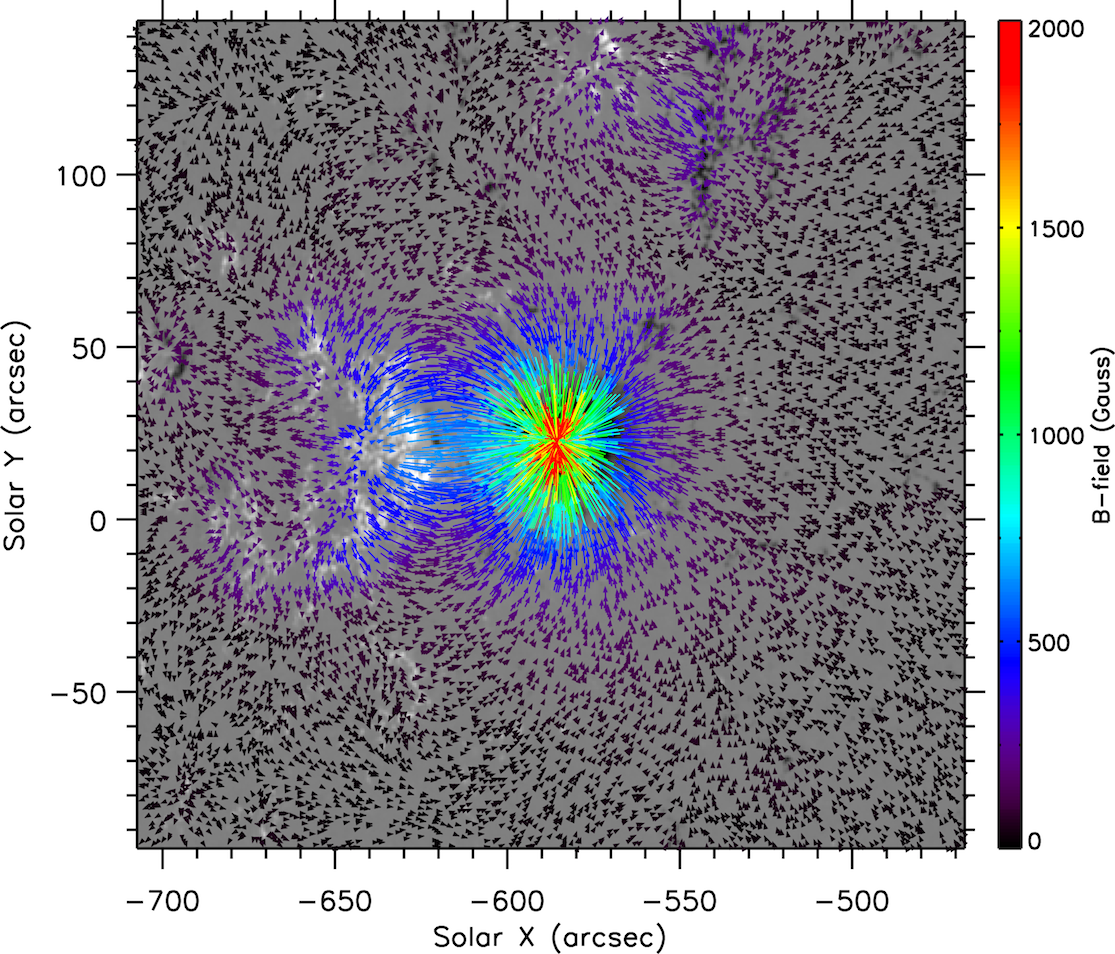}
\includegraphics[width=0.95\columnwidth, clip=true]{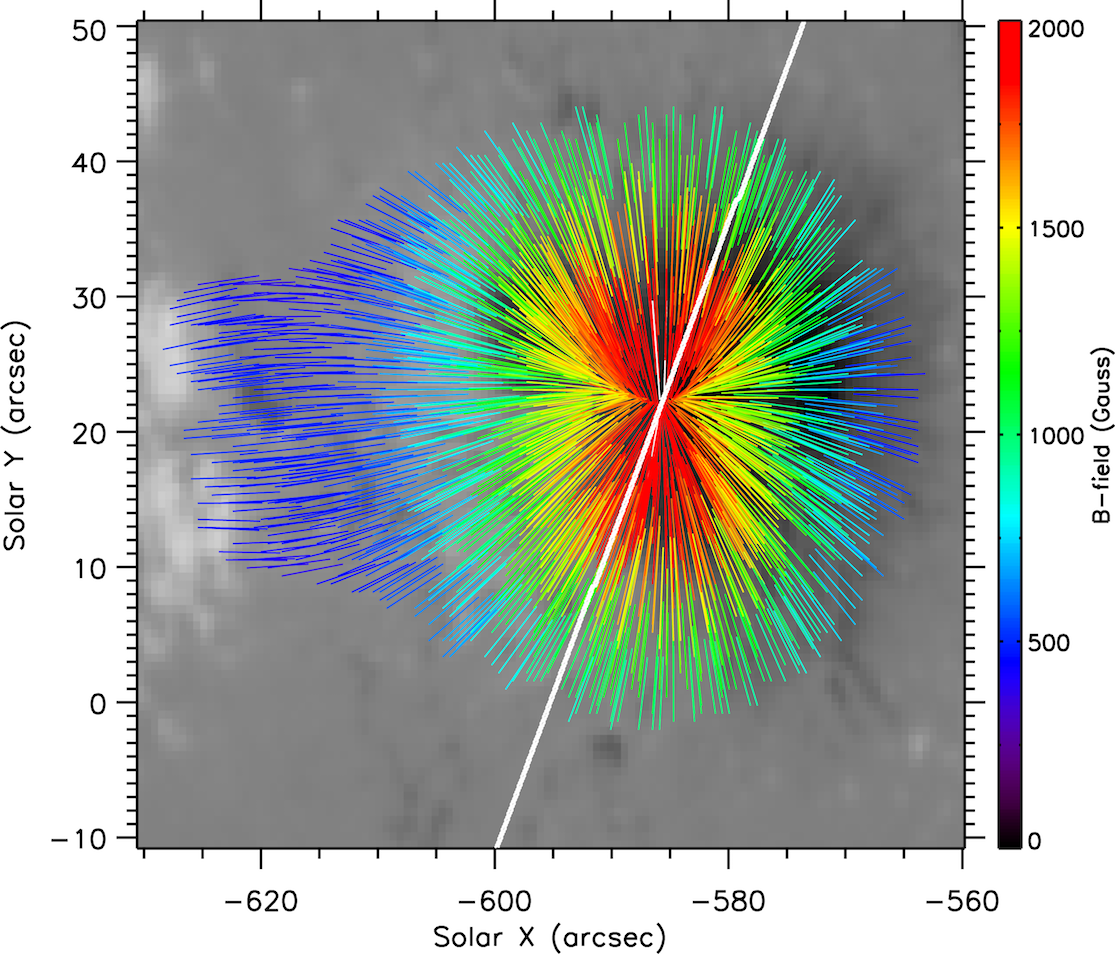}
\includegraphics[trim={0cm 0cm 0cm 0cm}, width=0.45\textwidth, clip=true]{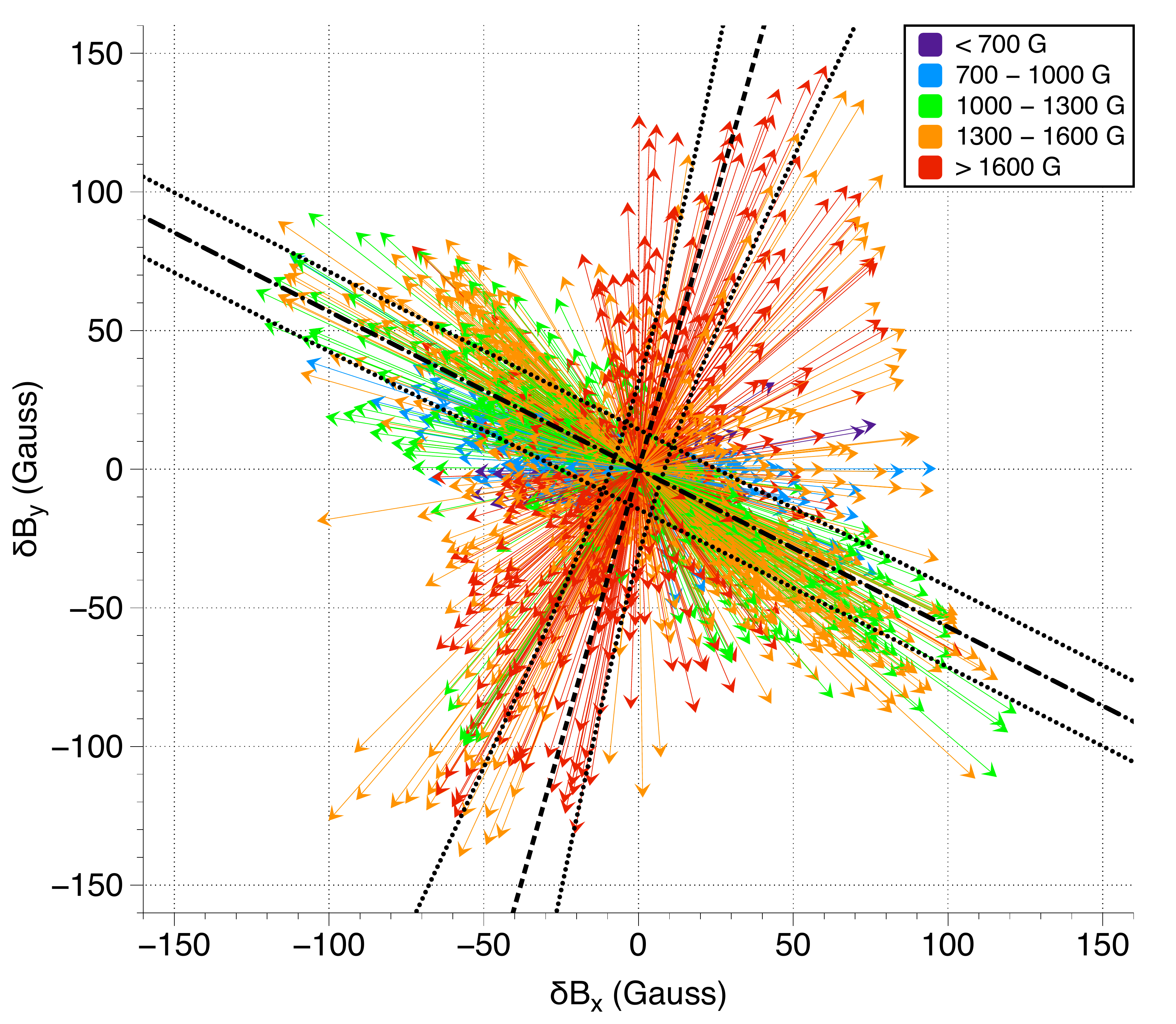}
\caption{{\it{Top Left Panel:}} An HMI photospheric vector magnetogram (black and white base image), overplotted with vector representations of the extrapolated chromospheric magnetic field. White and black colors in the base image represent positive and negative magnetic polarities, respectively, which have been saturated at $\pm1000$~G to aid visualization. The color scale corresponds to the absolute magnetic field strength of the depicted arrow vectors, while the lengths of the arrows relate to the magnitude of the transverse magnetic field. {\it{Top Right Panel:}} Zoom in of the extrapolations displayed in the upper-left panel, with the arrow heads removed for clarity. The white line represents the position of the FIRS slit across the center of the umbra. {\it{Bottom Panel:}} Vectorized representations of the changes in $B_{\mathrm{x}}$ and $B_{\mathrm{y}}$ obtained from UF pixels within the umbral region of the FIRS slit displayed in the top right panel. The color scale represents the absolute magnetic field strength for each arrow. The black dashed and dash-dotted lines highlight the lines of best fit for total magnetic fields exceeding and less-than-or-equal-to $1600$~G, respectively. For each line of best fit, the dotted black lines represent the associated $1\sigma$ uncertainties.}
\label{fig:Magnetogram}
\end{figure*}
%%%%%%%%%%%%%%%%%%%%%%%%%%%%%%%%%%%%%%%
%%%%%%%%%%%%%%%%%%%%%%%%%%%%%%%%%%%%%%%

 \subsection{Transverse Magnetic Field Perturbations}{\label{sec:Bx}}
The ability to accurately synthesize the observed Stokes $Q$ and $U$ spectra with HAZEL (see, e.g., Figure~{\ref{fig:Spectra}}) allows us to perform the first examination of the transverse component fluctuations of the magnetic field during UF events. Through use of the magnetic field strengths, $B$, in addition to the inclination and azimuthal angles, $\theta_{B}$ and $\chi_{B}$, respectively, we employed the techniques outlined by \citet{1990SoPh..126...21G} to decompose the parameters into orthogonal magnetic field components, $B_{\mathrm{x}}$ and $B_{\mathrm{y}}$, perpendicular to the solar normal. Here, $B_{\mathrm{x}}$ represents the magnetic field component running in the solar east-west direction, while $B_{\mathrm{y}}$ depicts the vector magnetic field orientated along the solar north-south axis. 

For each UF event registered, the un-signed measurements displayed in Figure~{\ref{fig:Histogram}} were converted into corresponding vector fluctuations, $\delta{B_{\mathrm{x}}}$ and $\delta{B_{\mathrm{y}}}$, perpendicular to the solar normal. These fluctuations are displayed in the bottom panel of Figure~{\ref{fig:Magnetogram}}. Here, the length of the arrow, which originates at the origin, represents the magnitude of the induced $B_{\mathrm{trans}}$ fluctuation, while the color scale corresponds to the total magnetic field strength, $B$, associated with that particular perturbation. It can be seen that the strongest magnetic fields (i.e., the red arrows in the bottom panel of Figure~{\ref{fig:Magnetogram}}, corresponding to $B > 1600$~G) are associated with the largest $B_{\mathrm{trans}}$ fluctuations. Furthermore, the $\delta{B_{\mathrm{x}}}$ and $\delta{B_{\mathrm{y}}}$ vector field perturbations associated with the largest magnetic fields are preferentially orientated along the north-north-west to south-south-east direction (i.e., $\strongarrow$), as indicated by the dashed trend line in the bottom panel of Figure~{\ref{fig:Magnetogram}}. This is in stark contrast to the weaker magnetic fields (i.e., $B \leq 1600$~G), where the associated $\delta{B_{\mathrm{x}}}$ and $\delta{B_{\mathrm{y}}}$ vector field fluctuations are more strictly orientated along the east-north-east to west-south-west direction (i.e, $\weakarrow$), as highlighted by the dash-dotted trend line in the bottom panel of Figure~{\ref{fig:Magnetogram}}. The black dotted lines also present in the bottom panel of Figure~{\ref{fig:Magnetogram}} represent the $1\sigma$ error boundaries of the fitted lines. Therefore, it is clearly evident that perturbations in the transverse vector magnetic field for weaker magnetic concentrations (i.e., $B \leq 1600$~G) are preferentially orientated in the east-north-east to west-south-west direction (i.e., $\weakarrow$), while the transverse vector magnetic field fluctuations associated with magnetic concentrations exceeding $1600$~G are preferentially orientated in the north-north-west to south-south-east direction (i.e., $\strongarrow$).

To investigate the cause of the preferential magnetic field deflection directions, we employed the non-linear force-free field (NLFFF) extrapolation code of \citet{2008JGRA..113.3S02W} to examine the widespread geometry of the magnetic field in the surrounding chromosphere. Since the vector magnetic field, derived from the FIRS {\HeI} spectrographic observations, only covers a $1{\,}.{\!\!}{\arcsec}125$ wide slot through the center of the sunspot umbra, it cannot be used to provide a more global overview of the surrounding magnetic field geometries. As a result, the scattered-light corrected HMI photospheric vector magnetograms were used for the NLFFF extrapolations. With the formation height of the {\HeI} spectral line known to reach an atmospheric ceiling of $\sim$2100~km \citep{1994IAUS..154...35A}, a two-dimensional cut-out of the resulting extrapolated magnetic fields, corresponding to the same atmospheric height of $\sim$2100~km, was isolated for subsequent study. 

A vector diagram of the extracted chromospheric magnetic fields is displayed in the top left panel of Figure~{\ref{fig:Magnetogram}}. Here, the arrows represent the positive-to-negative direction of the magnetic field, where their length corresponds to the magnitude of the transverse field component, $B_{\mathrm{trans}}$, and their color represents the total magnetic field strength, $B$, at the location displayed. It must be stressed that the arrow lengths and colors have been extracted from the three-dimensional magnetic field extrapolations of the active region, and are therefore independent from the HAZEL inversions of the FIRS spectro-polarimetric observations. A number of important features can be identified in the top left panel of Figure~{\ref{fig:Magnetogram}}. Firstly, the strongest magnetic fields (i.e., $B > 1600$~G), found towards the center of the sunspot umbra, display relatively short arrows; a consequence of $B_{\mathrm{z}} \gg B_{\mathrm{trans}}$. Secondly, for these strongest magnetic field concentrations, the corresponding transverse field vectors (i.e., $B_{\mathrm{x}}$ and $B_{\mathrm{y}}$ components) indicate a preferential north-north-west to south-south-east direction (i.e., $\strongarrow$). Thirdly, the weaker magnetic fields (i.e., $B \leq 1600$~G) entering the sunspot umbra have more extended arrow lengths, which is a consequence of their transverse magnetic field components being more dominant (i.e., $B_{\mathrm{trans}} \gtrsim B_{\mathrm{z}}$. Finally, the orientation of these weaker umbral magnetic fields is preferentially in the east-north-east to west-south-west direction (i.e., $\weakarrow$). This appears to be a consequence of the opposite polarity (i.e., positive) magnetic field concentration located immediately east of the sunspot. Here, the weaker magnetic fields originating within the sunspot will experience more rapid curvature towards the opposite polarity region; a consequence of the increased $B_{\mathrm{trans}} / B$ ratio within these locations where the plasma can be considered force-free \citep[i.e., where the plasma-$\beta < 1;$][]{2005LRSP....2....7L}. 

The top right panel of Figure~{\ref{fig:Magnetogram}} displays a zoom in to the sunspot umbra and its immediate surroundings. In an identical format to the top left panel of Figure~{\ref{fig:Magnetogram}}, the vector magnetic field components anchored into the sunspot umbra (also including those embedded in the neighboring positive polarity region) are displayed as colored lines, where the line color represents the total magnetic field strength, $B$, and the length of the line corresponds to the magnitude of the transverse field component, $B_{\mathrm{trans}}$. For ease of viewing, the arrow heads present in the top left panel of Figure~{\ref{fig:Magnetogram}} have been removed. The solid white line overplotted in the top right panel of Figure~{\ref{fig:Magnetogram}} represents the orientation of the FIRS slit, which was placed at an angle of $23.3{\degr}$ to the solar north-south axis. The preferential direction of the transverse magnetic field fluctuations (represented by the dashed line in the bottom panel of Figure~{\ref{fig:Magnetogram}}) associated with the strongest magnetic concentrations ($B > 1600$~G) is at an angle of $\approx$$14{\degr}$ to the solar north-south axis. Similarly, the transverse deflections associated with the weaker background magnetic fields ($B \leq 1600$~G) demonstrate a preferential angle of $\approx$$122{\degr}$ to the solar north-south axis (highlighted by the dash-dotted line in the bottom panel of Figure~{\ref{fig:Magnetogram}}). Since the preferential transverse magnetic field fluctuations are not exactly parallel or perpendicular to the slit orientation, this suggests that such characteristic deflection angles for the strong and weak magnetic fields are unlikely to be a purely systematic effect introduced during the data calibration process \citep[e.g., residual stray light contributions;][]{2007SPIE.6744E..1LZ, 2011A&A...535A.129B}. Furthermore, as the FIRS slit was placed through the center of the umbra, and therefore encapsulates magnetic fields spanning a plethora of azimuthal orientations, it is natural to expect the preferential directions of the transverse magnetic field fluctuations (bottom panel of Figure~{\ref{fig:Magnetogram}}) to correlate directly with the pre-existing field geometries revealed by the magnetic extrapolations in the top panels of Figure~{\ref{fig:Magnetogram}}. However, as the respective angles ($23.3{\degr}$ for the slit orientation compared with $\approx$$14{\degr}$ for the strong magnetic field fluctuations) are relatively close to one another, future data acquisition procedures may choose to implement a form of rotating slit assembly, similar in concept to that employed by the Reuven Ramaty High-Energy Solar Spectroscopic Imager \citep[RHESSI;][]{2002SoPh..210....3L}, to constantly evolve the slit orientation angle, hence minimizing and mitigating potential systematic effects.

Importantly, the top panels of Figure~{\ref{fig:Magnetogram}} depict the `at rest' quiescent geometry of the chromospheric magnetic field, while the bottom panel of Figure~{\ref{fig:Magnetogram}} reveals the dynamic fluctuations experienced by the two-dimensional transverse components of the magnetic field (i.e., $B_{\mathrm{x}}$ and $B_{\mathrm{y}}$) during UF shock phenomena. It can be clearly seen from Figure~{\ref{fig:Magnetogram}} that strong magnetic fields (i.e., $B > 1600$~G), which are normally orientated in the north-north-west to south-south-east direction (i.e., $\strongarrow$), experience transverse deflections along the same path following shock formation. Similarly, weaker magnetic fields (i.e., $B \leq 1600$~G) that are preferentially orientated east-north-east to west-south-west (i.e., $\weakarrow$), also demonstrate transverse deflections along the same direction once co-spatial UFs develop. Such characteristics can be related to a number of physical mechanisms. Firstly, an increase in the magnetic field inclination angle, $\theta_{B}$, would boost the transverse magnitude of the magnetic field (i.e., $B_{\mathrm{trans}}$) along the same two-dimensional direction, hence requiring no additional changes to the localized environment. Secondly, an increase in the measured magnetic field strength (i.e., as experienced by pixels where the magnetic flux tube expands into the umbral background throughout the shock event; `edge' pixels) along the direction of the vector field would also increase both components of the field perpendicular to the solar normal. Finally, the weaker fields embedded within the east-north-east to west-south-west (i.e., $\weakarrow$) geometries will have reduced magnetic tension, and therefore may be more susceptible to directional changes caused by the developing shock fronts. Of course, a combination of all three mechanisms may also contribute to the observed magnetic field fluctuations perpendicular to the solar normal. It must be noted that even though changes in the azimuthal angle, $\chi_{B}$, of the vector magnetic field may increase up to a maximum of $\sim$8 degrees following the creation of an UF, such a deflection is relatively minor, and as such, will not have a large impact on the vectorized plots displayed in Figure~{\ref{fig:Magnetogram}}. This further substantiates why the UF-induced changes in the transverse magnetic field components, $\delta{B_{\mathrm{x}}}$ and $\delta{B_{\mathrm{y}}}$, remain predominantly along the quiescent magnetic field vectors.

\section{Conclusions}{\label{sec:Conc}}
High resolution spectro-polarimetry and spectral imaging data has been employed in conjunction with advanced inversion techniques to provide a unique glimpse into magnetic field fluctuations in sunspot umbrae as a result of UFs. Through the comprehensive analysis of a large set (93{\,}991) of {\HeI} Stokes profiles, we find that a scatter diagram of temperature and magnetic field strength fluctuations provides evidence for two distinct populations. We uncover that the two populations relate to whether the detected signals originate at either the edge or center of the identified UF event. Edges of the UFs provide a positive correlation between magnetic field strength and temperature fluctuations, which is caused by the adiabatic expansion of the supporting magnetic flux tube into-and-through these pixel locations, hence simultaneously increasing the local magnetic field strength as the plasma is heated due to the dissipation of the shock front. This relationship allows us to derive the adiabatic index, $\gamma = 1.12\pm0.01$, for umbral locations in the lower solar atmosphere. Conversely, central pixels of the UFs provide an anti-correlation between the magnetic field strength and temperature perturbations, which is a result of the adiabatic expansion of the magnetic flux tube, causing a net magnetic field flux decrease in these central locations while the local plasma is being simultaneously heated through non-linear shock dissipation, providing credence to the adiabatic hypothesis put forward by \citet{2017ApJ...845..102H}. 

We have also shown, for the first time, fluctuations in the transverse components of the magnetic field (i.e., $B_{\mathrm{x}}$ and $B_{\mathrm{y}}$) resulting from UF phenomena. Through comparison with NLFFF extrapolations of scattered-light corrected HMI vector magnetograms, we find a number of possible scenarios to explain the observed transverse field perturbations: (1) changes in the inclination angles of the vector magnetic field, (2) increases in the measured magnetic field strength caused by the physical adiabatic expansion of the magnetic flux tube through those pixels, (3) reduced magnetic tension in the locations of weaker magnetic fields, thus promoting more susceptibility to field deflections, and (4) a combination of all three. Future work will require a close examination of all derived plasma parameters in order to address which scenario is most plausible. This will be a challenging task, requiring the segregation of weak/strong UFs, those in the rise/decay phases of their morphology, and those that may be forming over a range of optical depths and atmospheric heights. As such, it may be necessary to combine a multitude of complementary inversion routines, including NICOLE, HAZEL and the CAlcium Inversion using a Spectral ARchive \citep[CAISAR;][]{2015ApJ...798..100B, 2015A&A...582A.104R}, alongside a wealth of multi-wavelength spectro-polarimetric data, to further answer this question.

Greater spatial and temporal resolution, alongside higher polarimetric precision, would allow smaller-scale magnetic field fluctuations to be uncovered. Thankfully, new advanced observing facilities will soon be available to the community, including the Daniel K. Inouye Solar Telescope \citep[DKIST;][]{2004SPIE.5489..625K}, the National Large Telescope \citep[NLST;][]{2010SPIE.7733E..0IH} and the European Solar Telescope \citep[EST;][]{2013MmSAI..84..379C}. Such new facilities will likely allow us to reveal yet more information about the behavior of magneto-acoustic shock phenomena than current observing suites.   

%{\setlength{\parindent}{0cm}
%{\textit{Acknowledgements.}} 
\acknowledgements
S.J.H. thanks the Northern Ireland Department for the Economy for the award of a PhD studentship. D.B.J. wishes to thank the UK Science and Technology Facilities Council (STFC) for the award of an Ernest Rutherford Fellowship alongside a dedicated Research Grant. D.B.J. and S.D.T.G also wish to thank Invest NI and Randox Laboratories Ltd. for the award of a Research \& Development Grant (059RDEN-1) that allowed this work to be undertaken. A.A.R wishes to acknowledge financial support by the Spanish Ministry of Economy and Competitiveness through project AYA2014-60476-P. S.K.P. is grateful to the STFC for continued funding. The DST is operated by the National Solar Observatory (NSO). The NSO is operated by the Association of Universities for Research in Astronomy under cooperative agreement with the National Science Foundation. The SDO/HMI magnetograms employed in this work are courtesy of NASA/SDO and the AIA, EVE, and HMI science teams. %}

\newpage
%%\bibliographystyle{apj.bst}
%\bibliographystyle{aasjournal.bst}
%\bibliography{shouston}{}

\end{document}